\documentclass[aps,prb,twocolumn,showpacs,amsmath]{revtex4}
\usepackage{bm}
\usepackage{graphicx}

\newcommand{\mkright}{draft (\today)} 

\begin{document}

\title{Kinetic impedance and depairing in thin and narrow  superconducting films}

\author{John R. Clem and V. G. Kogan}
\affiliation{%
	Ames Laboratory--DOE and Department of Physics and Astronomy, 
	Iowa State University, Ames Iowa 50011, USA}

\date{\today}

\begin{abstract}
We use both Eilenberger-Usadel and Ginzburg-Landau (GL) theory to calculate the superfluid's temperature-dependent kinetic inductance  for all currents up to the depairing current in thin and narrow superconducting films.  The calculations apply to BCS weak-coupling superconductors with isotropic gaps and transport mean-free paths much less than the BCS coherence length. The kinetic inductance is calculated for the response to a small alternating current when the film is carrying a dc bias current. In the slow-experiment/fast-relaxation limit, in which the superconducting order parameter quasistatically follows the time-dependent current, the kinetic inductance diverges as the bias current approaches the depairing value.  However, in the fast-experiment/slow-relaxiation limit, in which the the superconducting order parameter remains fixed at a value corresponding to the dc bias current, the kinetic inductance rises to a finite value at the depairing current.  We then use time-dependent GL theory to calculate the kinetic impedance of the superfluid, which includes not only the kinetic reactance but also the kinetic resistance of the superfluid arising from dissipation due to order-parameter relaxation.  The kinetic resistance is largest for angular frequencies $\omega$ obeying   $\omega \tau_s > 1$, where $\tau_s$ is the order-parameter relaxation time, and for bias currents close to the depairing current.  We also include the normal fluid's contribution to dissipation in deriving an expression for the total kinetic impedance.  The Appendices contain many details about the temperature-dependent behavior of superconductors carrying current up to the depairing value.

\end{abstract}

\pacs{74.78.-w,74.78.Na,74.25.F-}%

\maketitle

\section{Introduction
\label{Sec_Intro}}

The kinetic inductance, arising chiefly from the kinetic energy of the superfluid, plays an important role in superconducting devices fabricated using thin and narrow superconducting films.\cite{Goltsman01,Annunziata10b,Natarajan12}  In such cases the kinetic inductance is generally much larger than the geometric inductance arising from stored magnetic energy.\cite{Meservey69,Clem05,Yoshida92} For example, the kinetic inductance plays a prominent role in determining the reset time of superconducting  single-photon detectors (SSPDs) fabricated with meandering superconducting lines.\cite{Hadfield05,Kerman06,Kerman07} Various calculations of the kinetic inductance, relevant to the performance of microstrip resonators\cite{Dahm97} and microwave kinetic inductance
detectors (MKIDs),\cite{Day03}  have been carried out using (a) the London equations neglecting the current-induced suppression of the order parameter,\cite{Meservey69,Clem05,Yoshida92,Brandt04} (b) the Ginzburg-Landau (GL) equations,\cite{Anlage89,Enpuku95,Annunziata10b} (c) a GL-inspired London-equation approach accounting  for the current-induced suppression of the order parameter, \cite{Bulaevskii11,Bulaevskii12} and (d) the BCS theory.\cite{Dahm97,Annunziata10b} Our goal in this paper is to present  theoretical calculations of the kinetic inductance for all temperatures and for all currents up to the depairing current for sample dimensions and properties applicable to present experimental studies of SSPDs\cite{Hadfield05,Hadfield09} and micro-resonators.\cite{Annunziata10b} Because these studies have used thin high-resistance films of  NbN,\cite{Goltsman01,Hadfield05,Kerman06,Kerman07,Annunziata10b,Bartolf10} Nb,\cite{Annunziata10b} NbTiN,\cite{Hortensius12} and TaN\cite{Engel12a,Engel12b} in the dirty limit, we adopt an isotropic s-wave BCS description, although many of  our results can be extended to apply under more general assumptions.

We consider  thin ($d \ll \lambda_0$) superconducting films of width $W$  much less than the two-dimensional screening length (Pearl length\cite{Pearl64}) $\Lambda = 2 \lambda_0^2/d$, where $\lambda_0$ is the temperature-dependent weak-field London penetration depth and $d$ is the film thickness.  The condition  $W \ll \Lambda$ guarantees that the self-field generated by the current has a negligible effect upon the current density $\bm j$, which therefore flows with the same spatial distribution as in the normal state.\cite{Clem11}  Moreover,  this condition also guarantees that the  inductance $L = L_m + L_{k}$ is dominated by the kinetic inductance of the superfluid $L_{k}$, which is typically larger than the geometric inductance $L_m$ (associated with the energy stored in the magnetic field) by a factor of order $\Lambda/W$.\cite{Clem05}  We focus on the calculation of the superfluid's kinetic inductivity ${\cal L}_{k}$. For a long strip of length $\ell$, width $W$, and thickness $d$, the kinetic inductance is $L_{k} = {\cal L}_{k} \ell/Wd$.

When the superconductor carries such a low current  that the superconducting order parameter is not significantly suppressed, the electromagnetic behavior is well described by the London equation, and the kinetic energy density of the superfluid can be expressed as\cite{Meservey69}
\begin{equation}
U_k = \frac{1}{2}n_{s0} m v_s^2 = \frac{1}{2}\Big(\frac{m}{n_{s0}e^2}\Big)j_s^2 = \frac{1}{2}\mu_0 \lambda_{0}^2 j_s^2 = \frac{1}{2}{\cal L}_{k0}j_s^2,
\label{uk}
\end{equation}
where $n_{s0}$ is the superfluid density, $m$  the electron mass, $v_s$ the  superfluid velocity component in the $x$ direction, $j_s =-n_{s0} e v_s$  the supercurrent density  component in the $x$ direction, and $-e$  the electron charge,  such that the kinetic inductivity of the superfluid obeys\cite{Tinkham96,foot1}
\begin{equation}
{\cal L}_{k0}(T) =\mu_0 \lambda_{0}^2(T)= m/n_{s0}e^2.
\label{Lks0}
\end{equation}
The subscripts $0$ on $n_{s0}$, $\lambda_{0}$, and  ${\cal L}_{k0}$ are a reminder that these quantities apply in  the limit as $j_s \to 0$.   The simple relationship given in Eq.\ (\ref{Lks0}) has made possible the determination of  $\lambda_0(T)$ vs $T$ in YBa$_2$Cu$_3$O$_{7-\delta}$ from kinetic-inductance measurements.\cite{Lee93}

When the superconductor carries high currents, however, calculation of the superfluid's kinetic inductance becomes more complicated, especially when the current density approaches the depairing value $j_d$. In high currents it is no longer  possible to define the kinetic inductance by considering only the stored kinetic energy density as in Eq.\ (\ref{uk}), because, as examined in detail in Appendix \ref{EnergySec}, increasing $j_s$ to large values  also affects the superconducting condensation energy by suppressing the superconducting order parameter.  To account for this effect, we take advantage of  Maki's\cite{Maki63a,Maki63b,Maki64}  recognition that the current-induced suppression of the order parameter in a thin film can be treated using a pair-breaking parameter exactly analogous to that used by Abrikosov and Gor'kov\cite{AG61} in their study of the effect of paramagnetic impurities upon superconductivity.  The fusion of the theories for these two problems\cite{Maki65} has resulted in a large body of related work,\cite{Skalski64,Ambegaokar65,Nam67a,Nam67b,Nam70,Maki69} much of which we summarize for the benefit of the reader in  Sec.\ \ref{jsSec} and the Appendices.  As noted by previous authors,\cite{Kupr80,Romijn82,Anthore03,Semenov08,Groll10} a convenient starting point for this purpose is the mean-field Eilenberger-Usadel theory.\cite{Eilenberger68,Usadel70}

An additional complication in calculating the kinetic inductance is that $n_s$, which  depends on  the superconducting order parameter,  can change only on a time scale slower than a variety of difficult-to-determine internal relaxation times,\cite{Anlage89,Tinkham96,Kopnin01,Kaplan76} which we here represent crudely by a single relaxation time $\tau_s$.  As a consequence, $n_s$ may or may not be able to follow the changes in $j_s$ and $v_s$.

Calculations of the current dependence of ${\cal L}_{k}$ are simplified in two limiting cases:\cite{Anlage89}  (a) slow experiments (fast relaxation, Sec.\ \ref{SlowSec}), in which $j_{s}$ and $v_s$ vary on an experimental time scale $\tau_{exp}$  much longer than the relaxation time $\tau_s$, such that the order parameter and the superfluid density $n_s$ quasistatically follow $j_{s}$ and $v_s$, and (b) fast experiments (slow relaxation, Sec.\ \ref{FastSec}),  in which  $j_s$ and $v_s$ change so rapidly about their time averages $\bar j_{s}$  and $\bar v_{s}$ (on a time scale $\tau_{exp}$ much shorter than $\tau_s$) that the order parameter and $n_s$ cannot track the time dependence, and $n_s$ remains very close to the value corresponding to  $\bar j_{s}$ and $\bar v_s$.  To provide an approximation to the transition between these two limiting cases, in Sec.\ \ref{Complex} we employ a simplified phenomenological model based on the time-dependent GL (TDGL) equations\cite{Tinkham96} to calculate the complex impedivity due to the superfluid.  In Sec.\ \ref{Normal} we include the  normal-fluid's resistive contribution to the total complex impedivity, and in Sec.\ \ref{discussion} we provide a brief summary and discussion of our results.  Various details of the calculation are included in Appendices 
A-D.

\section{Superfluid-velocity dependence of the supercurrent density and the depairing current density \label{jsSec}}

The purpose of this section is to explain clearly how to calculate the many effects of the current-induced suppression of the order parameter that have been obtained by previous authors.  We need a formalism that allows us to  calculate the depairing current density $j_d(T)$ in superconductors with a short normal-state mean-free path at all temperatures in the superconducting state.  A compact way of doing this is to employ the  quasiclassical Eilenberger\cite{Eilenberger68} theory as formulated by Usadel\cite{Usadel70} for the dirty limit.

 Consider a superconducting strip  extending along the $x$ direction when the current is uniform. Let $j_s$ and $A_s= mv_s/e $ denote the $x$ components of the supercurrent density $\bm j_s$ and  the gauge-invariant vector potential $\bm A_s = \bm A + (\phi_0/2\pi) \nabla \gamma$, where $\bm A$ is the gauge-dependent vector potential, $\phi_0 = h/2e$ the superconducting flux quantum, and $\gamma$ the gauge-dependent phase of the superconducting order parameter.  

The  supercurrent density can always be expressed as $j_s = -n_s e v_s$, but in general $n_s$ is a function of the superfluid velocity $v_s$ and has the value $n_{s0}$ when $v_s = 0$ but decreases monotonically to zero as $|v_s|$ increases. 
For positive $j_s$ and negative values of $v_s$, the supercurrent density $j_s = n_s e |v_s|$ initially increases linearly as a function of $|v_s|$, reaches a maximum $j_d(T)$ (the depairing or pair-breaking current density) at $|v_s| = v_d(T)$, then decreases to zero  at $|v_s| = v_m(T)$. 

When the superconductor is current-biased, only the portion of the curve $j_s$ vs $|v_s|$ for $0 \le |v_s| \le v_d(T)$ is accessible.  On the other hand, following a suggestion by Fulde and Ferrell,\cite{Fulde63} Bhatnagar and Stern\cite{Bhatnagar68,Bhatnagar73} showed that it is possible to probe experimentally the shape of  $j_s$ vs $|v_s|$ even for $v_d(T) \le |v_s| \le v_m(T)$ using a multiply-connected sample geometry.  In this paper we first examine the behavior of  $j_s$ over the full range of values of $v_s$, but later in applying these results to study the kinetic inductance we limit our attention to  the current-biased case in which $j_s$ is a single-valued function of $v_s$ in the range $0 \le j_s \le j_d$.

\subsection{Depairing current density calculated from the Usadel equations\label{UsadelSec}}

For the problem at hand the Usadel equations can be written as\cite{Usadel70}
\begin{eqnarray}
&& -\hbar D(GF^\prime -FG^\prime)^\prime=2\Delta G -2\hbar\omega_n F\,,\label{U1}\\
&&G^2+|F|^2=1\,,\label{U2}\\
&& \Delta\,\ln\frac{T_{c0}}{T}= 2\pi k_BT 
 \sum_{n=0}^{\infty}\left(\frac{\Delta}{\hbar 
\omega_n}-F\right)\,, \label{U3}\\
&&j_s = -4\pi e N(0)Dk_BT\sum_{n=0}^{\infty} {\rm Im}F^*F^\prime\,,
\label{U4}
\end{eqnarray}
where $\Delta$ is the superconducting order parameter, $D = v_F^2 \tau/3 = v_F \ell/3$ is the diffusivity, $v_F$ is the average velocity of electrons at the Fermi surface, $\tau$ is the normal-state transport lifetime, $\ell$ is the mean-free path, $\hbar \omega_n = (2n+1)\pi k_B T$ is the Matsubara frequency, $N(0)$ is the density of Bloch states of one spin at the Fermi level, $T$ is the temperature, and $T_{c0}$ is the zero-current transition temperature. The primes in Eq.\ (\ref{U1}) denote differentiation with respect to $x$.   These equations describe supercurrent flow in a superconductor with an s-wave isotropic gap in the weak-coupling limit of the BCS theory.\cite{BCS57}  

However, this mean-field theory does not account for  the possibility that one- or two-dimensional fluctuations could grow to produce phase slips or vortex crossings.  

Since $W \ll \Lambda$, we can neglect the self-field of the current and  choose a gauge for which we may replace the gauge-invariant gradient $\bm\nabla +2\pi i \bm A/\phi_0$ by $\hat x \partial_x$. 
Looking for solutions  
 of the form
$\Delta =\Delta_q e^{iqx}$, $F=F_{nq} e^{iqx}$, and $G=G_{nq}$, where $q$ is the gradient of the phase of the order parameter, we find that Eqs.\ (\ref{U1}) and (\ref{U2}) become
\begin{eqnarray}
&&QF_{nq}G_{nq}=\Delta_q G_{n q}-\hbar\omega_n F_{nq}\,,
\label{U1a}\\
&&\quad G_{n q}^2+F_{n q}^2=1\,,
\label{U2a}
\end{eqnarray}
where $Q = \hbar D q^2/2$.

\begin{figure}
\includegraphics[width=8cm]{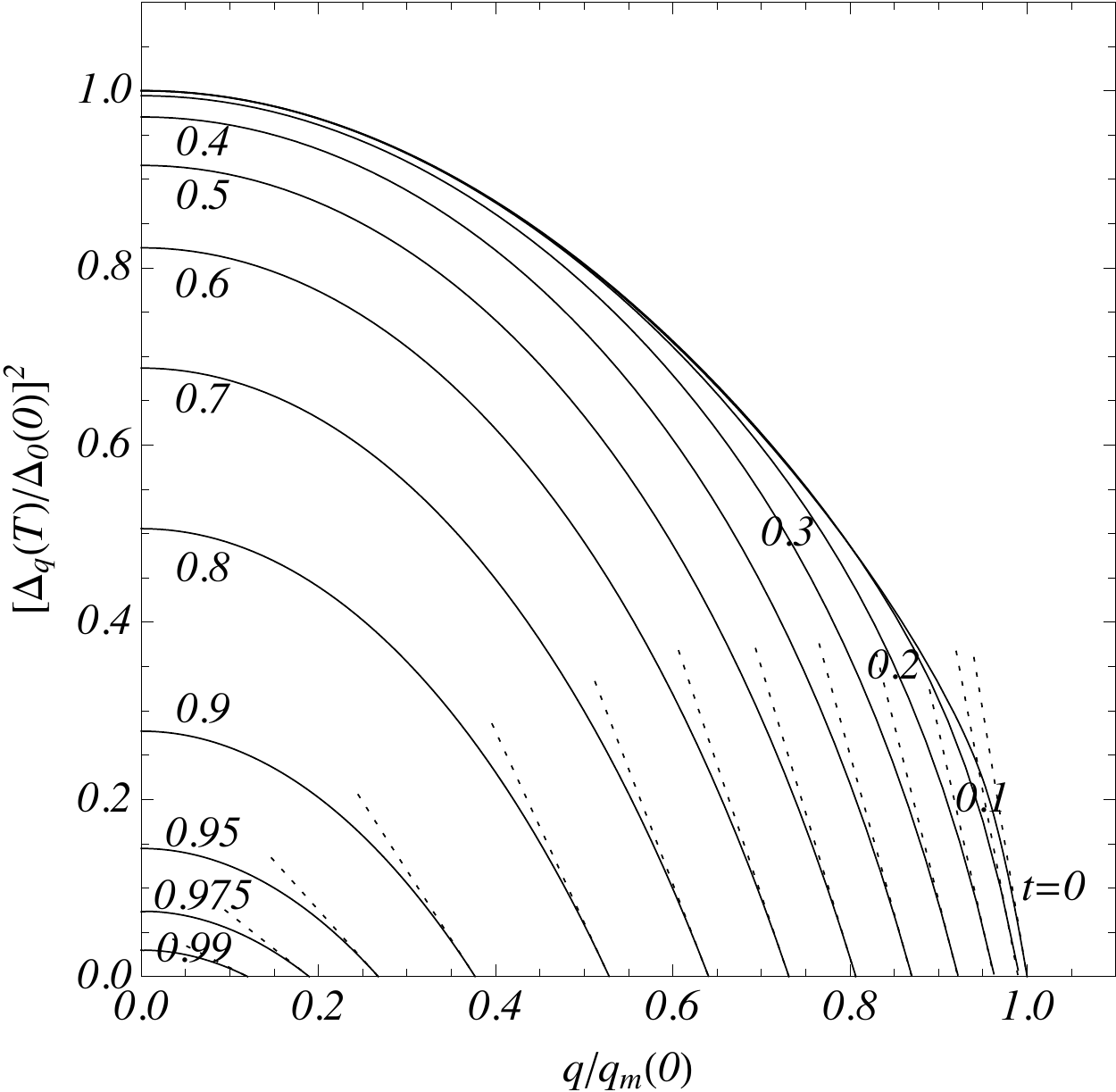}
\caption{%
$[\Delta_q(T)/\Delta_0(0)]^2$ vs $q/q_m(0)$ obtained from Eqs.\ (\ref{selfcons}) and (\ref{unqsoln}) for various values of $t = T/T_{c0}$. The dotted lines show the linear behavior as $[\Delta_q(T)/\Delta_0(0)]^2 \to 0$ in the limit as $q \to q_m(T)$. }
\label{Deltaqtfig}
\end{figure}

\begin{figure}
\includegraphics[width=8cm]{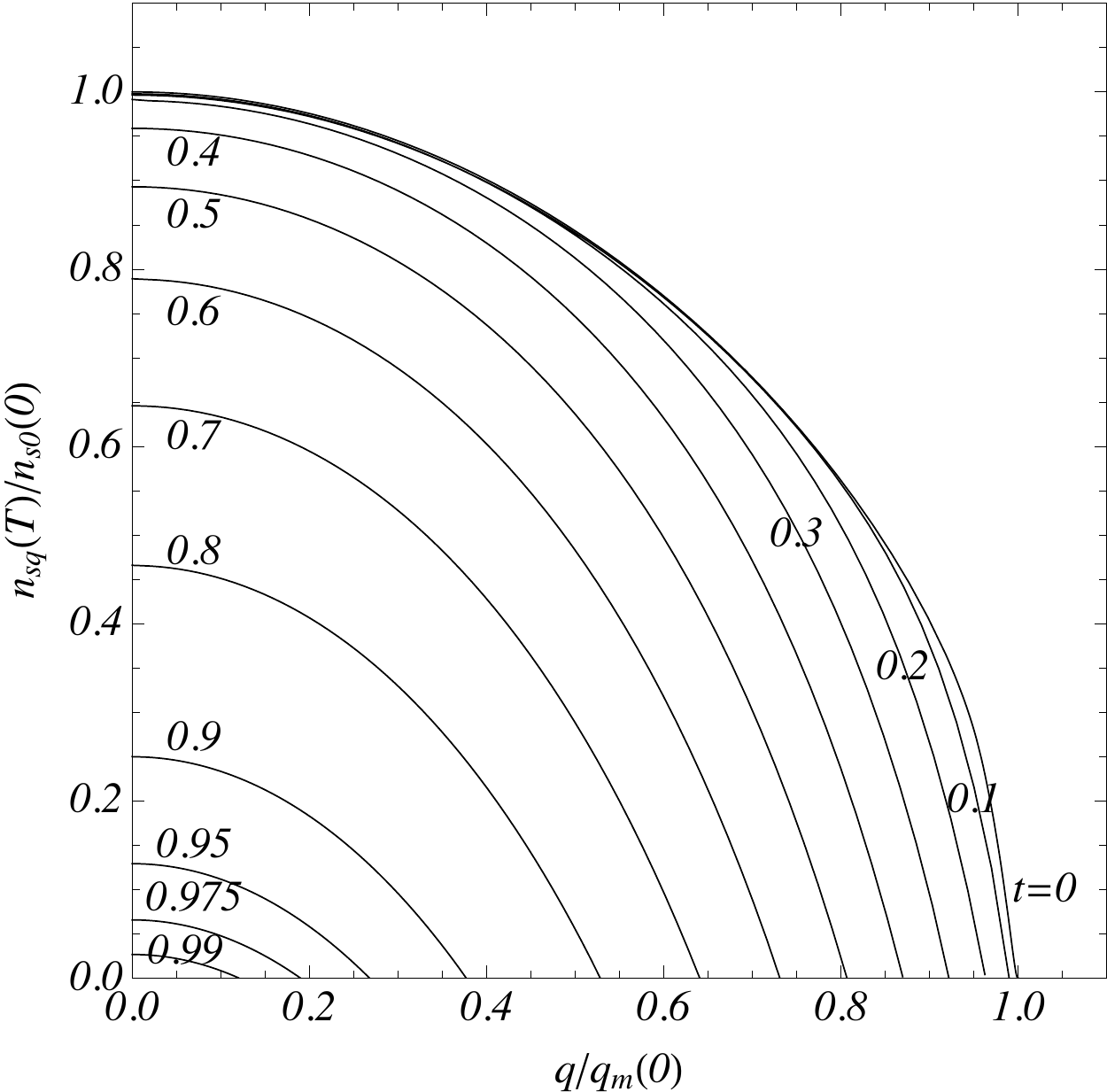}
\caption{%
$n_{sq}(T)/n_{s0}(0)=\lambda_0^2(0)/\lambda_q^2(T)$ vs $q/q_m(0)$ obtained from Eqs.\ (\ref{selfcons}), (\ref{nsqsum}), and (\ref{unqsoln}) for various values of $t = T/T_{c0}$. $n_{sq}(T)/n_{s0}(0) \to 0$ as $q \to q_m(T)$.}
\label{nsqtfig}
\end{figure}

Throughout this paper, the symbol $q$ appears frequently; it can be regarded as a compact abbreviation for the gauge-invariant  vector potential $A_s$, the superfluid velocity $v_x$, or the gradient of the phase $\gamma$ of the order parameter, since all these quantities are related via $q = 2\pi A_s/\phi_0 = 2mv_s/\hbar = \gamma/x.$  The presence of a subscript $q$ indicates that the subscripted quantity is a function of $q$.  We show later that as $q$ increases, the pair-breaking effect reduces the superconducting transition temperature $T_{cq}$, the order parameter $\Delta_q$ (Figs.\ \ref{Deltaqtfig} and \ref{Deltaqfig}) and the superfluid density $n_{sq}$ (Fig.\ \ref{nsqtfig}), and increases the penetration depth $\lambda_q$ [Eq.\ (\ref{nsq&lambdaq})].  We show here that these $q$ dependencies combine to produce the current dependence of the kinetic inductance shown in Figs.\ \ref{invLksqfig} and \ref{Lksqffig}.  However, the $q$ dependence of $\lambda_q$ could be shown somewhat more directly in sensitive measurements of the penetration depth in the Meissner state as a function of applied ac and dc magnetic fields.

As noted by Maki,\cite{Maki69} Eqs.\ (\ref{U1a}) and (\ref{U2a}) are equivalent to those of the Abrikosov-Gor'kov (AG) theory\cite{AG61}  for pair-breaking scattering, except for the replacement of the AG spin-flip scattering rate $1/\tau_m$ by $Dq^2/2$.  Our results therefore share many properties with those of the AG theory.  For example, we show that the transition temperature $T_{cq}$ depends upon $q$ and decreases monotonically from its value $T_{c0}$ at $q = 0$ to zero at a critical value of  $q$ given by  $q_m(0)=1/\xi(0)=(\pi\xi_0\ell /3)^{-1/2}$, where $\xi_0 = \hbar v_F/\pi \Delta_0(0)$ is the BCS coherence length.\cite{BCS57} For a fixed value of $q$, the order parameter $\Delta_q(T)$ is nonzero only for temperatures $T$ less than $T_{cq}$; equivalently, for a fixed temperature $T$, the order parameter $\Delta_q(T)$ is nonzero only for values of $q$ less than $q_m(T)=1/\xi(T)$. 

Introducing $u_{nq} = G_{nq}/F_{nq},$  we find that Eqs.\ (\ref{U1a}) and (\ref{U2a}) can be written as  
\begin{equation}
\frac{\eta}{\epsilon} = u_{nq}\left
(1-\frac{\zeta}{ \sqrt{1+u_{nq}^2}}\right), 
\label{unqomegaq}
\end{equation}
and
\begin{equation}
F_{nq}= \frac{1}{ \sqrt{1+u_{nq}^2}},
\label{AGF}
\end{equation}
where $\eta = n+1/2$, $\epsilon = \Delta_q/2\pi k_B T$ and $\zeta = Q/\Delta_q$.
$u_{nq}$ (which  depends implicity upon $T$)  can be obtained for arbitrary values of $\eta$, $\epsilon$, and $\zeta$ by solving Eq.\ (\ref{unqomegaq}) as a quartic equation [see Appendix \ref{unq}].     

With the introduction of $u_{nq}$, two equations remain to be solved.  The self-consistency equation (\ref{U3}) in the presence of the current becomes, for general values of $\Delta_q$, $\omega$, $Q$, and $T$,
\begin{equation} 
  \ln \frac{1}{t}= \sum_{n=0}^{\infty}\left(\frac{1}{n+1/2}- \frac{1 } { \epsilon\sqrt{1+u_{nq}^2}}\right).\, 
\label{selfcons}
\end{equation}
In general, $\Delta_q(T)$ must be obtained by numerically solving Eq.\ (\ref{selfcons}) using Eq.\  (\ref{unqsoln}), but the results can be checked against analytic results  obtainable in the limits of $q \to 0$ and $q \to q_m(T)$.   Figure \ref{Deltaqtfig} shows $[\Delta_q(T)/\Delta_0(0)]^2$ as a function of $q/q_m(0)$ for a series of values of the reduced temperature $t = T/T_{c0}$. (Closely related plots were given as Fig.\ 1 in Ref.\ \onlinecite{Skalski64} and Fig.\ 4 in Ref.\ \onlinecite{Ambegaokar65}.)  The sums for $t \ge 0.1$ were evaluated by summing $n$ from 0 to 500, but  for $t = 0$ we used the analytic results in Eqs.\ (\ref{deltaless})-(\ref{zetadef}).

From the  current equation (\ref{U4}) we find that when the superfluid velocity $v_s$ is in the $x$ direction, the supercurrent density in that direction is\cite{foot1}
\begin{equation}
j_{sq}(T) = -n_{sq}(T) e v_s.
\label{jx}
\end{equation}
From Eqs.\ (\ref{U4}), (\ref{AGF}) and (\ref{jx}) we obtain a general expression for the  $q$-dependent superfluid density,
\begin{equation}
n_{sq}(T) = \frac{8\pi m N(0)Dk_BT}{\hbar}
\sum_{n=0}^\infty\frac{1}{1+u_{nq}^2}.
\label{nsqT}
\end{equation}
When $q \to 0$, we have $u_{n0} = 2\pi k_BT(n+1/2)/\Delta_0(T)$  [see Eq.\ (\ref{unqomegaq})], and when this is used in Eq.\ (\ref{nsqT}), evaluation of the sum  yields
\begin{equation}
n_{s0}(T) = \frac{2\pi m N(0)D \Delta_0(T)}{\hbar}
\tanh\Big(\frac{\Delta_0(T)}{2k_B T}\Big),
\label{ns0T}
\end{equation}
such that 
\begin{equation}
n_{s0}(0) = \frac{2\pi m N(0)D \Delta_0(0)}{\hbar}
\label{ns0TA}
\end{equation}  
and
\begin{equation}
\frac{n_{sq}(T)}{n_{s0}(0)} = \frac{4k_BT}{\Delta_0(0)}
\sum_{n=0}^\infty\frac{1}{1+u_{nq}^2}.
\label{nsqsum}
\end{equation}

In general, $n_{sq}(T)/n_{s0}(0)$  must be obtained by numerically solving Eqs.\ (\ref{selfcons}) and  (\ref{nsqsum}) using Eq.\ (\ref{unqsoln}).   Figure \ref{nsqtfig} shows $n_{sq}(T)/n_{s0}(0)$  as a function of $q/q_m(0)$ for a series of values of the reduced temperature $t = T/T_{c0}$.  The sums for $t \ge 0.1$ were evaluated by summing $n$ from 0 to 500, but  for $t = 0$ we used the analytic results in Eqs.\ (\ref{nsq01})-(\ref{nsq02}).

As shown in Fig.\ \ref{nsqtfig}, $n_{sq}(T)/n_{s0}(0)$ depends upon $q$ and vanishes at $q = q_m(T)$ (see Appendix \ref{Deltato0}).  The corresponding $q$-dependent penetration depth $\lambda_q(T)$ can be obtained from 
\begin{equation}
\frac{n_{sq}(T)}{n_{s0}(0)}=\frac{\lambda_0^2(0)}{\lambda_q^2(T)}.
\label{nsq&lambdaq}
\end{equation}
Note, however, that current-biased experiments can access  values of $q$ only up to $q_d(T)$, where the magnitude of the current density reaches the depairing limit $j_d(T)$. 

The general expression for the supercurrent density is
\begin{equation}
j_{sq}(T) =-\frac{n_{sq}(T)e^2A_s}{m}= -\frac{A_s}{\mu_0 \lambda_q^2(T)}.
\label{jsq}
\end{equation}  
From Eq.\ (\ref{jx}) or (\ref{jsq}) we see that, because $j_{sq}(T)$ is the product of $n_{sq}(T)$ (a monotonically decreasing function of $q$) and $ev_s = e^2A_s/m = e\hbar q/2m$, the magnitude of $j_{sq}(T)$ reaches a maximum, called the depairing current density $j_d(T)$, when $q = q_d(T)$, where $0 < q_d(T) < q_m(T)$.  We define $\tilde j_q(T)$ as the magnitude of $j_{sq}(T)$   normalized to $n_{s0}(0)ev_m(0)=\phi_0/2\pi\mu_0\lambda_0(0)^2\xi(0)$, such that 
\begin{equation}
\tilde j_q(T) = \frac{n_{sq}(T)}{n_{s0}(0)} \frac{q}{q_m(0)}.
\label{tildejq}
\end{equation}
The maximum value of $\tilde j_q(T)$ vs $q$ is the normalized depairing current density $\tilde j_d(T)$.  

\begin{figure}
\includegraphics[width=8cm]{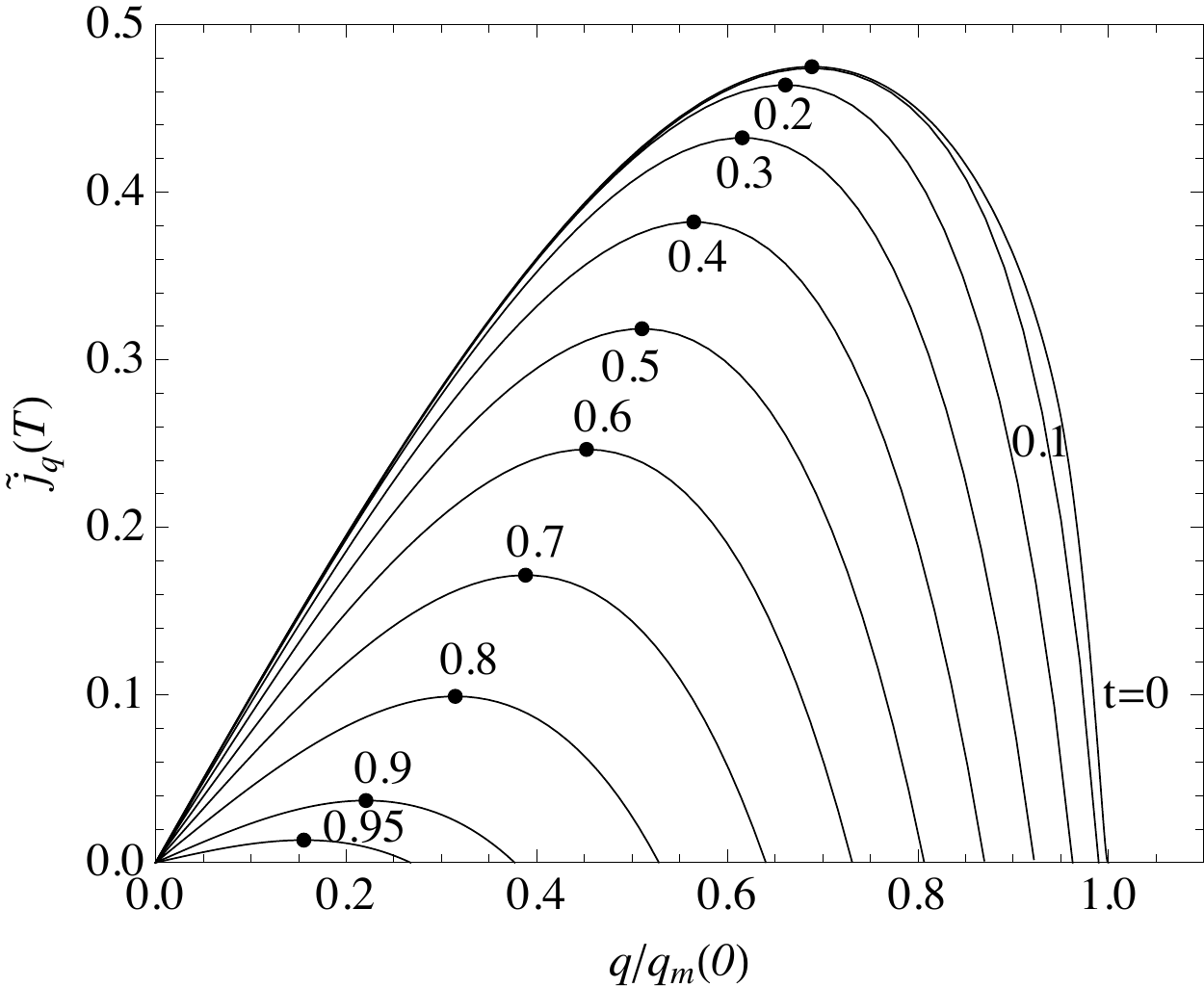}
\caption{%
$\tilde j_q(T)$ vs $q/q_m(0)$ obtained from Eqs.\ (\ref{selfcons}), (\ref{nsqsum}),  (\ref{tildejq}), and (\ref{unqsoln}) for various valus of  $t = T/T_{c0}$. The points label the values of $\tilde j_q$ and $q$ corresponding to the depairing current density $j_d$ and $q_d$. Current-biased experiments probe only the portions of the curves to the left of these points.  $\tilde j_q(T) \to 0$ as $q \to q_m(T)$.}
\label{tildejqtfig}
\end{figure}

\begin{figure}
\includegraphics[width=8cm]{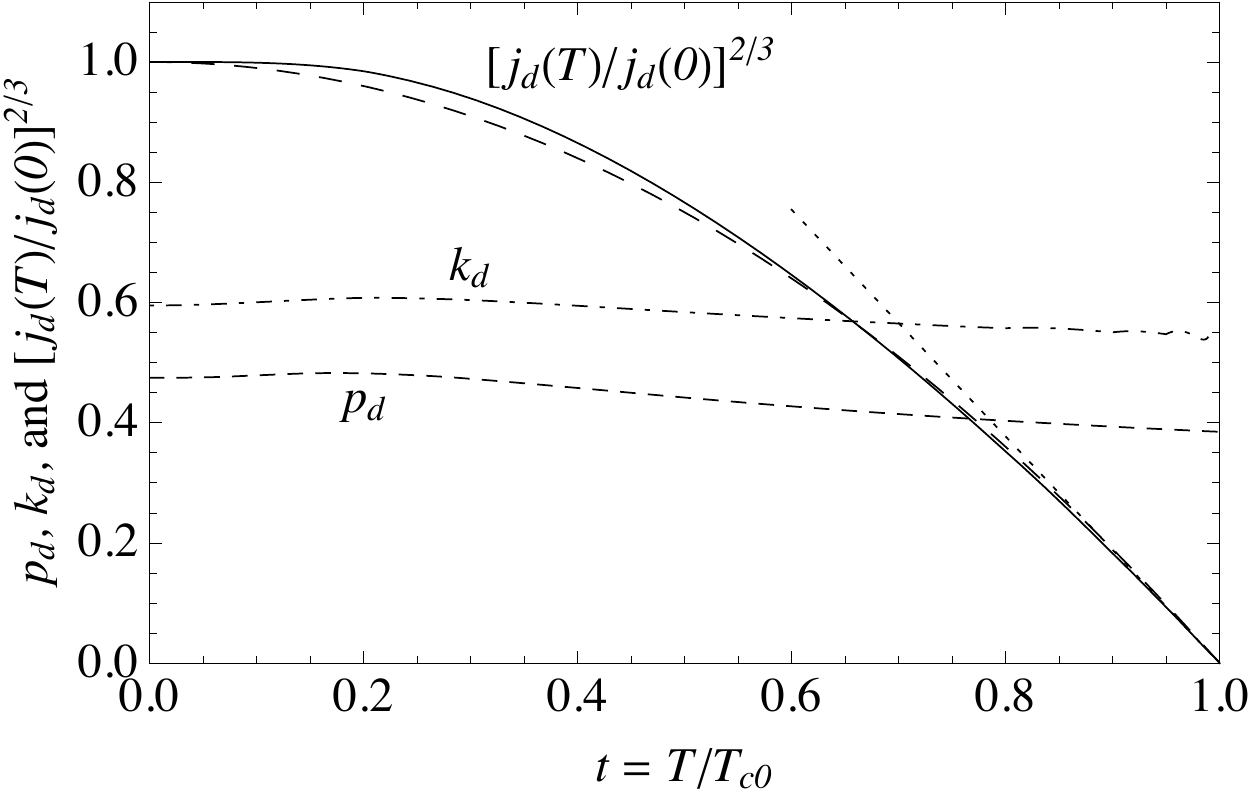}
\caption{%
$[j_d(T)/j_d(0)]^{2/3}$ (solid) vs $t = T/T_{c0}$ obtained numerically from Eqs.\ (\ref{selfcons}), (\ref{nsqsum}),  (\ref{tildejq}), and (\ref{unqsoln}). The dotted line shows the behavior of $j_d(T)$ in the GL limit near $T_{c0}$ [Eq.\ (\ref{jdGLtau})], and the long-dashed curve shows the approximation\cite{Bardeen62,Romijn82,Kunchur04} $j_d(T)/j_d(0)\approx (1-t^2)^{3/2}$.  The short-dashed curve shows the variation of $p_d(T)$ [Eq.\ (\ref{jdTpd})] from 0.475 at $t = 0$ to 0.385 at $t = 1$, and the dot-dashed curve shows the variation of $k_d(T)$ [Eq.\ (\ref{jdTkd})]  from 0.595 at $t = 0$ to 0.544 at $t = 1$. }
\label{jdTwoThirdsfig}
\end{figure}

 In general, $\tilde j_q(T)$  must be obtained numerically from Eqs.\ (\ref{selfcons}), (\ref{nsqsum}),  (\ref{tildejq}), and (\ref{unqsoln}), but the results can be checked against analytic results  obtainable in the limits $t \to 0$ and $t \to 1$, to be discussed in more detail later in Secs.\ \ref{T0} and \ref{GLjd}.  
Figure \ref{tildejqtfig} shows the general behavior of $\tilde j_q(T)$  as a function of $q/q_m(0)$ for a series of values of the reduced temperature $t = T/T_{c0}$. (Plots similar to Fig.\ \ref{tildejqtfig} were shown as Fig.\ 1 in Ref.\ \onlinecite{Maki63b} and Fig.\ 2 in Ref.\ \onlinecite{Romijn82}.) The points label the values of $\tilde j_q$ and $q$ corresponding to the depairing current density $j_d$ and $q_d$.    
The solid curve in Fig.\ \ref{jdTwoThirdsfig} shows $[j_d(T)/j_d(0)]^{2/3}=[\tilde j_d(T)/\tilde j_d(0)]^{2/3}$   as a function of $t = T/T_{c0}$, and the dotted line illustrates how $j_c(T)$ approaches the $(1-t)^{2/3}$ behavior in the GL regime close to $T_{c0}$.  

For all temperatures below $T_{c0}$, estimates of $j_d(T)$ can be obtained from
\begin{equation}
j_d(T) = p_d(T) \frac{\phi_0}{2\pi\mu_0\lambda_0^2(T)\xi(T)},
\label{jdTpd}
\end{equation} 
where $p_d(T)$ is a dimensionless function defined by Eq.\ (\ref{jdTpd}). The dashed curve in Fig.\ \ref{jdTwoThirdsfig} shows $p_d(T)$, which is obtained numerically from all the other quantities in Eq.\ (\ref{jdTpd}),  varies from 0.475 at $T = 0$ [Eq.\ (\ref{jd0pd})] to 0.385 as $T \to T_{c0}$ [Eq.\ (\ref{jcTcpd})] with a maximum of 0.483 at $t = T/T_{c0} = 0.17$.

Similarly, estimates of the depairing current density also can be obtained for all temperatures from
\begin{equation}
j_d(T) = k_d(T) H_c(T)/\lambda_0(T),
\label{jdTkd}
\end{equation} 
where the dimensionless quantity $k_d(T)$, defined by Eq.\ (\ref{jdTkd}) and shown by the dot-dashed curve in Fig.\ \ref{jdTwoThirdsfig}, varies from 0.595 at $T = 0$ [Eq.\ (\ref{jd0pd})] to 0.544 as $T \to T_{c0}$ [Eq.\ (\ref{jcTcpd})] with a maximum of 0.608 at $t = T/T_{c0} = 0.21$.  The values of $k_d$ shown in Fig.\ \ref{jdTwoThirdsfig} were obtained from 
\begin{equation}
k_d(T)=0.595\frac{\tilde j_d(T)}{\tilde j_d(0)}\frac{H_c(0)}{H_c(T)}\frac{\lambda_0(T)}{\lambda_0(0)}
\end{equation}
via Eqs.\ (\ref{jd0H0}), (\ref{HcT}) and (\ref{ns0TByns00}), where $H_c(T)$ is the temperature-dependent bulk thermodynamic critical field (see Appendix \ref{EnergySec}).  Plots similar to Fig.\ \ref{jdTwoThirdsfig} were given as Fig.\ 9 in Ref.\ \onlinecite{Kupr80} and Fig.\ 4 in Ref.\ \onlinecite{Romijn82}.

\subsection{Depairing current density at zero temperature\label{T0}}

\begin{figure}
\includegraphics[width=8cm]{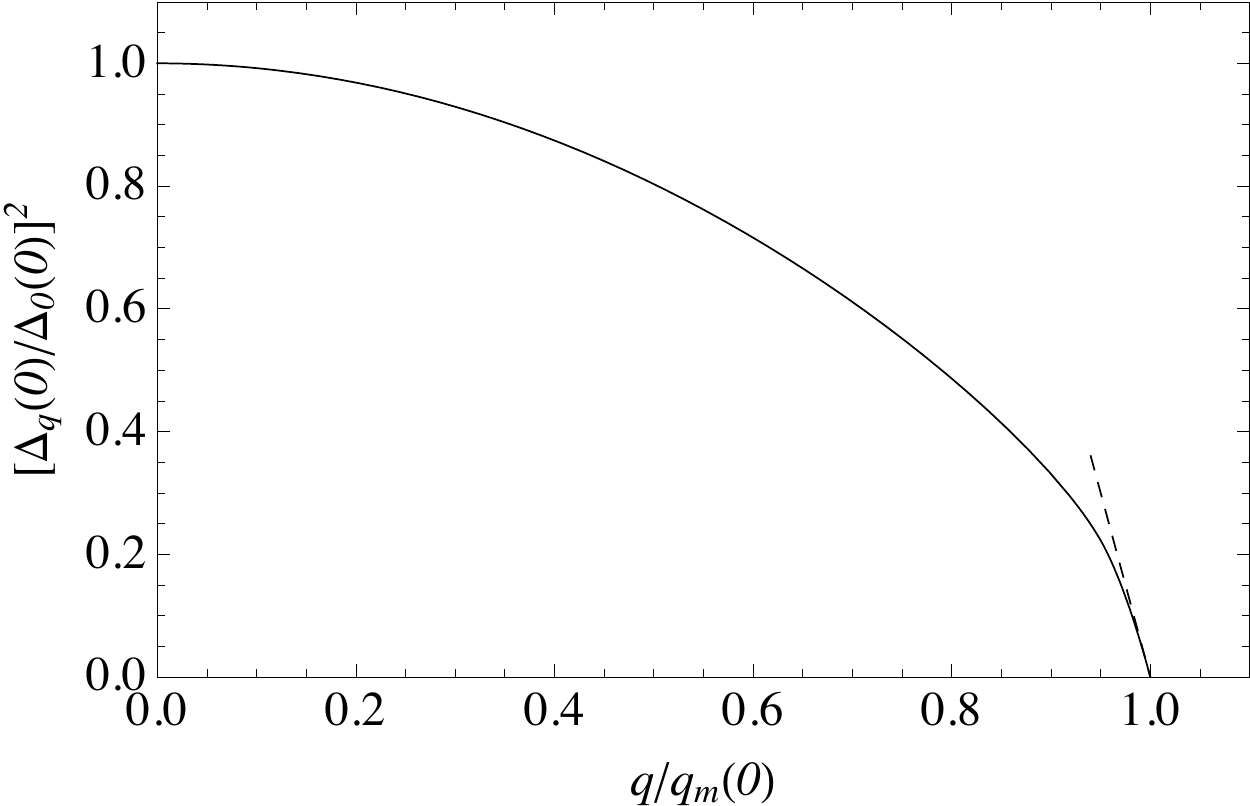}
\caption{%
$[\Delta_q(0)/\Delta_0(0)]^2$ vs $q/q_m(0)$ obtained from Eqs.\ (\ref{deltaless})-(\ref{zetadef}). The dashed line shows the linear behavior $6[1-q/q_m(0)]$ as $q \to q_m(0)$. }
\label{Deltaqfig}
\end{figure}

At $T = 0$, the $q$ dependence of $\Delta_q(0)$ can be obtained by converting the sum in Eq.\ (\ref{selfcons}) to an integral over $u_{nq}$.   The result is\cite{AG61,Maki69}
\begin{eqnarray}
\frac{\Delta_q(0)}{\Delta_0(0)}&=&\exp(-\pi \zeta_0/4), \;\;0 \le \zeta_0 \le 1,\label{deltaless}\\
&=&\exp[-\big(\zeta_0\sin^{-1}\zeta_0^{-1}-\sqrt{1-\zeta_0^{-2}}\big)/2 \nonumber\\
&&-\cosh^{-1}\zeta_0],\;\;\zeta_0 \ge 1,\label{deltamore}
\end{eqnarray}
where 
\begin{equation}
\zeta_0 = \frac{\hbar D q^2}{2\Delta_q(0)} = \frac{1}{2}\Big(\frac{q}{q_m(0)}\Big)^2 \frac{\Delta_0(0)}{\Delta_q(0)}.  
\label{zetadef}
\end{equation}
Figure \ref{Deltaqfig} shows $[\Delta_q(0)/\Delta_0(0)]^2$, obtained from numerical solution of Eqs.\ (\ref{deltaless})-(\ref{zetadef}), as a function of $q/q_m(0)=v_s/v_m(0)$. 

\begin{figure}
\includegraphics[width=8cm]{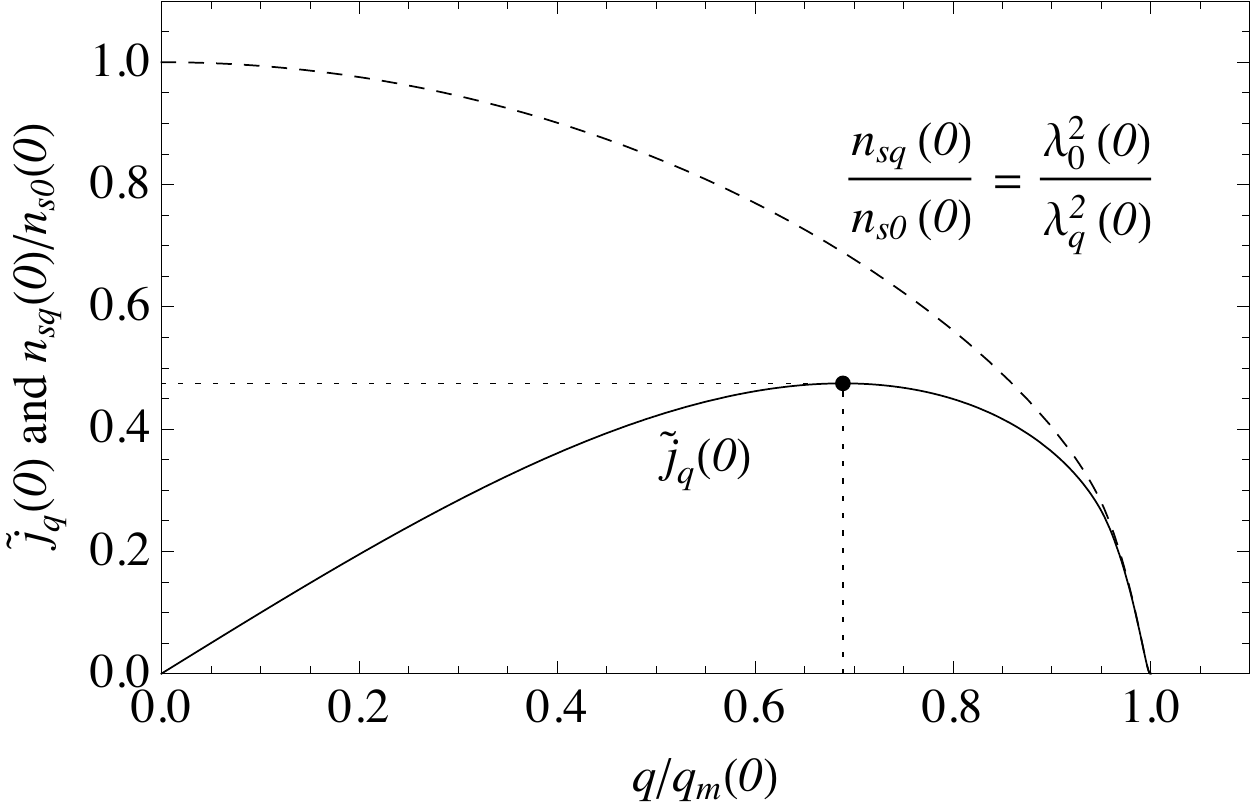}
\caption{%
  Reduced superfluid density $n_{sq}(0)/n_{s0}(0)=\lambda_0^2(0)/\lambda_q^2(0)$  (dashed  curve) vs $q/q_m(0) = v_s/v_m(0)$, obtained from Eqs.\ (\ref{nsq01}) and (\ref{nsq02}), and normalized $q$-dependent supercurrent density $\tilde j_q(0)$ (solid  curve) vs $q/q_m(0) = v_s/v_m(0)$, obtained from Eq.\ (\ref{jq0}).  The black point and dotted lines indicate the maximum $\tilde j_q(0)$, the normalized depairing supercurrent density $\tilde j_d = 0.475$, which occurs at $q_d/q_m(0) = 0.689$. }
\label{jq0fig}
\end{figure}

The $q$ dependence of $n_{sq}(0)$ can be obtained in a similar way.  The result is\cite{Maki69}
\begin{eqnarray}
\frac{n_{sq}(0)}{n_{s0}(0)}&=&\exp(-\pi \zeta_0/4)(1 -4\zeta_0/3\pi), \;\;0 \le\zeta_0 \le 1,\label{nsq01}\\
&=&\exp(-\pi \zeta_0/4)\{\frac{2}{3\pi \zeta_0^2}[(1+2\zeta_0^2)\sqrt{\zeta_0^2-1}-2\zeta_0^3]\nonumber\\
&&+\frac{2}{\pi}\sin^{-1}\zeta_0^{-1}\},\;\;\zeta_0 \ge 1,\label{nsq02}
\end{eqnarray}
where $\zeta_0$  is given by Eq.\ (\ref{zetadef}).  $n_{sq}(0)$ is shown as the dashed curve in Fig.\ \ref{jq0fig}.  The $q$-dependent penetration depth at zero temperature $\lambda_q(0)$ can be obtained from 
\begin{equation}
\frac{n_{sq}(0)}{n_{s0}(0)}=\frac{\lambda_0^2(0)}{\lambda_q^2(0)}.
\label{nsqbyns0}
\end{equation}

To obtain the depairing current density, consider the $q$-dependent (or $v_s$-dependent) supercurrent density   at $T = 0$ [Eq.\ (\ref{jx})], normalized to $-n_{s0}(0)ev_m(0)=-\phi_0/2\pi\mu_0\lambda_0(0)^2\xi(0)$,
\begin{equation}
\tilde j_q(0) = \frac{n_{sq}(0)}{n_{s0}(0)}\Big(\frac{q}{q_m(0)}\Big),
\label{jq0}
\end{equation}
shown as the solid curve in Fig.\ \ref{jq0fig}. (Plots similar to Fig.\ \ref{jq0fig} were shown as Fig.\ 1 in Ref.\ \onlinecite{Maki63b} and Fig.\ 2 in Ref.\ \onlinecite{Romijn82}.) The point and dotted lines show the  maximum $\tilde j_q(0)$, the normalized depairing supercurrent density $\tilde j_d(0) = 0.475$, which occurs at $q_d(0)/q_m(0) = 0.689$ and $\zeta_0 = 0.300$.\cite{Maki63b} The resulting zero-temperature depairing supercurrent density can be expressed in several  ways:
\begin{eqnarray}
j_d(0) &=& 0.475 \frac{\phi_0}{2\pi\mu_0\lambda_0^2(0)\xi(0)},
\label{jd0pd}\\
&=&1.491 N(0)e[\Delta_0(0)]^{3/2}\sqrt{D/\hbar},
\label{jd0Gorkov}\\
&=&0.595 H_c(0)/\lambda_0(0).
\label{jd0H0}
\end{eqnarray}
Equation (\ref{jd0Gorkov}) coincides with the result given for the depairing supercurrent density in Ref.\ \onlinecite{Kupr80}.  Here $H_c(T)$ is the temperature-dependent bulk thermodynamic critical field (see Appendix \ref{EnergySec}), and $H_c(0) = \Delta_0(0)\sqrt{N(0)/\mu_0}.$

\subsection{Depairing current density in the  GL regime\label{GLjd}}

As shown by Gor'kov\cite{Gorkov58}, the GL theory of superconductivity\cite{Ginzburg50} is derivable from the microscopic BCS theory\cite{BCS57} at temperatures $T$ very close to the transition temperature $T_{c0}$.  This assures us that  we also can apply the Usadel equations to recover the GL results.  We present the GL depairing-current results here for completeness, even though they are so well known that they appear in textbooks.\cite{Tinkham96}
For $T$ close to $T_{cq}$ or, equivalently, for $q$ close to $q_m(T)$, the $q$-dependent order parameter $\Delta_q(T)$ becomes  small, and it is useful to expand $u_{nq}$ and $1/\epsilon\sqrt{1+u_{nq}^2}$ in powers of $\epsilon = \Delta_q(T)/2\pi k_B T$:
\begin{eqnarray}
\frac{1}{\epsilon\sqrt{1+u_{nq}^2}}&=&\frac{1}{\eta+\alpha}-\frac{\epsilon^2}{2}\Big[\frac{1}{(\eta+\alpha)^3}-\frac{\alpha}{(\eta+\alpha)^4}\Big]\nonumber\\
&&+{\rm O}(\epsilon^4),
\label{uExpansion}
\end{eqnarray}
where 
\begin{equation}
\alpha = \frac{e^{-\gamma}}{4t}\frac{q_m^2(T)}{q_m^2(0)}= \frac{0.140}{t}\frac{q_m^2(T)}{q_m^2(0)}.
\label{alpham2}
\end{equation} 

Substituting this into Eq.\ (\ref{selfcons}) and keeping only the lowest order terms, since we know that  $\epsilon = \Delta_q(T)/2\pi k_B T \ll 1$ when $1-t \ll 1$, where $t = T/T_{c0}$, we obtain  
\begin{eqnarray} 
  \ln \frac{T_{c0}}{T}&=& \sum_{n=0}^{\infty}\Big\{\frac{1}{n\!+\!1/2}- \frac{1 } {n\!+\!1/2\! +\!\alpha} \nonumber\\
 &+&\!\!\Big[\frac{1 } {(n\!+\!1/2\! +\!\alpha)^3}-\frac{\alpha } {(n\!+\!1/2 \!+\!\alpha)^4}\Big]\frac{\epsilon^2}{2}\Big\},
\label{GLselfcons}
\end{eqnarray}
  The sums can be expressed in terms of  digamma functions and their derivatives. When $1-t \ll 1$, Eq.\ (\ref{GLselfcons}) has solutions only for $\alpha \ll 1$ and can be expanded as
\begin{equation}
1-t = \frac{\pi^2 e^{-\gamma}q^2}{8q_m^2(0)}+\frac{7\zeta(3) \epsilon^2}{2}.
\end{equation}
Solving for $\epsilon^2$, dividing by $\epsilon_0^2 = 2(1-t)/7\zeta(3)$, and making use of $q_m^2(T) =[8e^\gamma (1-t)/\pi^2]q_m^2(0)$ (see Appendix \ref{Deltato0}), we obtain
\begin{equation}
\frac{\Delta_q^2(T)}{\Delta_0^2(T)}=1-\frac{q^2}{q_m^2(T)},
\label{reducedDeltaSqGL}
\end{equation} 
where $q_m(T) = 1/\xi(T)$.

Substituting the expansion of Eq.\ (\ref{uExpansion}) into Eq.\ (\ref{nsqsum}), keeping only the lowest order terms, we obtain
\begin{equation}
\frac{n_{sq}(T)}{n_{s0}(T)}=\frac{\lambda_0^2(T)}{\lambda_q^2(T)}=\frac{\Delta_q^2(T)}{\Delta_0^2(T)}=f^2=1-\frac{q^2}{q_m^2(T)},
\label{nsqGL}
\end{equation}
and 
\begin{equation}
\frac{n_{sq}(T)}{n_{s0}(0)}=\frac{\lambda_0^2(0)}{\lambda_q^2(T)}=\frac{4 \pi e^\gamma}{7\zeta(3)}\Big(1-\frac{q^2}{q_m^2(T)}\Big)(1-t).
\end{equation} 
The reduced $q$-dependent supercurrent density becomes 
\begin{equation}
\tilde j_q(T)=\frac{4 \pi e^\gamma}{7\zeta(3)}\frac{q_m(T)}{q_m(0)}\Big(1-\frac{q^2}{q_m^2(T)}\Big)\frac{q}{q_m(T)}(1-t),
\label{tildejqGL}
\end{equation}
whose maximum occurs at $q_d(T)/q_m(T) = 1/\sqrt{3}$, such that (see Appendix \ref{Deltato0}) the reduced depairing current density is
\begin{equation}
\tilde j_d(T)=\frac{16 \sqrt{2} e^{3\gamma/2}}{21\sqrt{3}\zeta(3)}(1-t)^{3/2}=1.230 (1-t)^{3/2}.
\label{jdGLtau}
\end{equation} 
Thus, in the GL regime the depairing current density can be expressed as: 
\begin{equation}
j_d(T) = 0.385 \frac{\phi_0}{2\pi\mu_0\lambda_0^2(T)\xi(T)},
\label{jcTcpd}
\end{equation} 
where 0.385 =2/3$\sqrt{3}$, 
\begin{equation}
j_d(T) = 3.865N(0)e[\Delta_0(0)]^{3/2}\sqrt{D/\hbar}(1-t)^{3/2},
\label{jcTN0}
\end{equation}or, since $\sqrt{2}H_c = \phi_0/2\pi\mu_0\lambda_0\xi$ in the GL theory,
\begin{equation}
j_d(T) = 0.544 H_c(T)\lambda_0(T),
\label{jcTcHc}
\end{equation} 
where 0.544 = (2/3)$^{3/2}$.

To simplify calculations in  the GL limit later in Secs.\ \ref{SlowSec} and \ref{FastSec} we introduce the parameter $\phi$ ($0\le \phi \le \pi/2$), such that 
\begin{eqnarray}
|j_{sq}|/j_d &=& \sin\phi,\label{jsByjdGL}\\
q/q_m(T) &=& (2/\sqrt{3})\sin(\phi/3),\\
f^2&=&[1+2\cos(2\phi/3)]/3.\label{f2GL}
\end{eqnarray}

\section{Superfluid kinetic inductivity in slow experiments (fast relaxation) \label{SlowSec}}

In Sec.\ \ref{jsSec} we have summarized the results of previous authors and provided details of how to account quantitatively for the current-induced suppression of the order parameter in dirty thin-film superconductors at all temperatures in the superconducting state.
We are now in a position to calculate the kinetic inductivity of the superfluid measured in slow, low-frequency  (or, equivalently, fast relaxation) current-biased experiments, in which both $j_s$ [$|j_s| \le j_d(T)$] and $v_s$ [$|v_s| \le v_d(T)$] vary on a time scale $\tau_{exp}$ much longer than the relaxation time $\tau_s$ required for the superconducting order parameter to change.\cite{Anlage89,Tinkham96,Kopnin01}   

In a one-dimensional conductor carrying a uniform current the gauge-invariant electric potential $P = \Phi -(\phi_0/2\pi)d\gamma/dt$ is zero,\cite{foot2} and the electric field along the conductor is $E = -dA_s/dt={\cal L}_{k}(q,T)dj_{sq}(T)/dt$.\cite{foot3} 
Since from Eq.\ (\ref{jsq}) we have $A_s=  (\phi_0/2\pi)q=-j_{sq}(T)\mu_0 \lambda_q^2(T)$, taking the  time derivative and using  $df/dt = (df/dq) dq/dt$ we obtain the kinetic inductivity of the superfluid for slow experiments, 
\begin{eqnarray}
{\cal L}_{k}(q,T) &=& \mu_0\Big[\frac{d}{dq}\Big(\frac{q}{\lambda_q^2(T)}\Big)\Big]^{-1}=\Big|\frac{dj_{sq}(T)}{dA_s}\Big|^{-1}\nonumber \\
&=&\mu_0\lambda_0^2(T)F_{s}\Big(\frac{|j_s|}{j_{d}(T)}\Big),
\label{LksSlow}
\end{eqnarray}
where the slow-experiment function $F_s$ is simply ${\cal L}_{k}(q,T)/\mu_0\lambda_0^2(T)$ but expressed as a function of the normalized current density $|j_s|/j_{d}(T)$ rather than as a function of $q$.  
While $q$ is a convenient theoretical variable, $j_s$ is a more convenient variable for the current-biased case.  For $|q| < q_d(T)$ and $|j_s| < j_d(T)$, $j_{s}$ is a single-valued function of $q$, shown in Figs.\ \ref{tildejqtfig} and \ref{jq0fig}. 

\begin{figure}
\includegraphics[width=8cm]{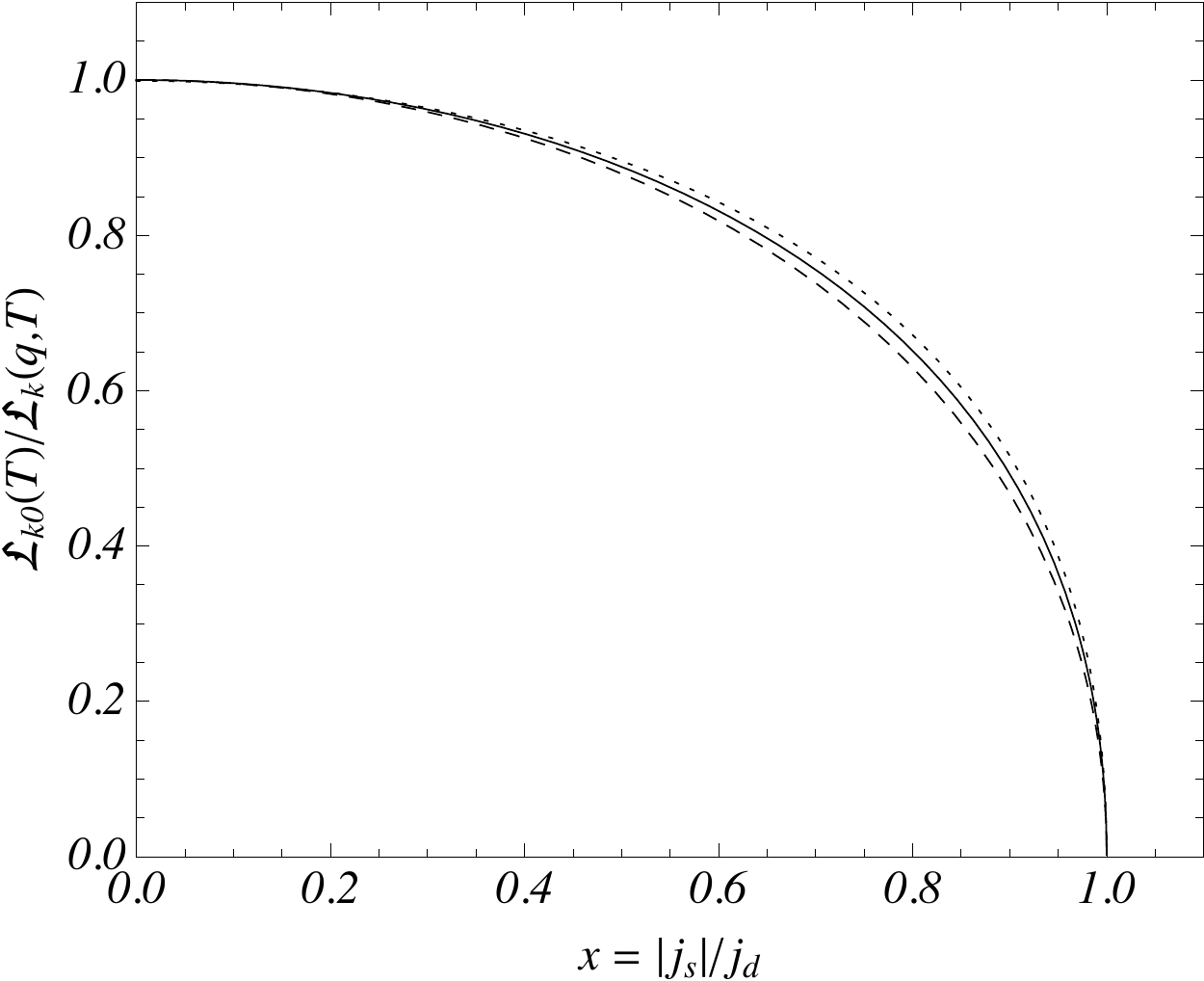}
\caption{%
${\cal L}_{k0}(T)/{\cal L}_{k}(q,T) = 1/F_{s}(|j_s|/j_{d}(T))$ for slow experiments vs $x=|j_s|/j_d(T)$ at $t = T/T_{c0} = 0$ (solid), $t =0.3$ (dotted) and $t \to 1$ (dashed). Note that ${\cal L}_{k0}(T) ={\cal L}_{k}(0,T)=\mu_0 \lambda_{0}^2(T)$.}
\label{invLksqfig}
\end{figure}

In the limit of small currents, when $q \to 0$, ${\cal L}_{k}(q,T)$ reduces to ${\cal L}_{k}(0,T)={\cal L}_{k0}(T)$ [Eq.\ (\ref{Lks0})].  However, as can be seen from Figs.\  \ref{tildejqtfig} and \ref{jq0fig}, $|dj_{sq}/dq|$ decreases monotonically for increasing values of $q$ and becomes zero at the depairing value.  Accordingly, as $q$ increases, ${\cal L}_{k}(q,T)$ starts from ${\cal L}_{k0}(T)$, increases monotonically, and diverges at $q = q_d(T)$, where $|j_s| = j_d(T)$. 
Because ${\cal L}_{k}(q,T)/{\cal L}_{k0}(T)$ diverges as $|j_s|\to j_d$, we show in Fig.\ \ref{invLksqfig} the typical dependence of the inverse, ${\cal L}_{k0}(T)/{\cal L}_{k}(q,T) = 1/F_{s}(|j_s|/j_{d}(T))$, vs $|j_s|/j_d(T)$.  This figure was obtained by (a)  evaluating $\tilde j_q(T)$ [Eq.\ \ref{tildejq})] and $d\tilde j_q(T)/dq$ numerically for $t = T/T_{c0}$ = 0, 0.1, 0.2, 0.3, 0.4, 0.5, 0.6, 0.7, 0.8, and 0.9, and analytically [see Eq.\ (\ref{tildejqGL})] in the GL limit $t\to 1$, (b) calculating  
\begin{equation}
\frac{{\cal L}_{k0}(T)}{{\cal L}_{k}(q,T)} = \frac{n_{s0}(0)}{n_{s0}(T)}q_m(0)\frac{d\tilde j(T)}{dq}
\end{equation}
 using Eq.\ (\ref{ns0T}) to evaluate $n_{s0}(0)/n_{s0}(T)$, and (c) making a parametric plot of ${\cal L}_{k0}(T)/{\cal L}_{k}(q,T)$ vs $|j_s|/j_d = \tilde j_q(T)/\tilde j_d(T)$.   As shown by the solid curve for $t = 0$, the dotted curve for $t = 0.3$, and the dashed curve for $t \to 1$, the behavior of ${\cal L}_{k0}(T)/{\cal L}_{k}(q,T)$ vs $|j_s|/j_d(T)$ is not monotonic as the temperature changes, but the curves for all other temperatures (not shown) lie in a narrow band between the dotted and dashed curves.  As $|j_s|/j_d \to 1$, all the curves have an inverse-square-root dependence close to that in the GL limit $t\to 1$,
\begin{equation}
{\cal L}_{k0}^{GL}(T)/{\cal L}_{k}^{GL}(q,T)=(2\sqrt{6}/3)[1-|j_s|/j_d(T)]^{1/2}.
\end{equation}

The  curves shown in  Fig.\ \ref{invLksqfig} can be represented  by $y_{sn}(x) = (1-x^n)^{1/n}$ (not shown in  Fig.\ \ref{invLksqfig}), which  fits the calculated values of $y = {\cal L}_{k0}(T)/{\cal L}_{k}(q,T)$ vs $x = |j_s| /j_d(T)$ for $0\le x < 0.97$  with 1\% accuracy  for $(n,t)$ = (2.21, 0), (2.21, 0.1), (2.27, 0.2), (2.30, 0.3), (2.28, 0.4), (2.25, 0.5), (2.22, 0.6), (2.18, 0.7), (2.16, 0.8), (2.13, 0.9), and (2.11, $t\to 1$).

To calculate the  kinetic inductivity in the GL limit shown by the dashed curve in Fig.\  \ref{invLksqfig}, it is convenient to use the parametric relations $x = |j_s|/j_d = \sin\phi$ and $y = {\cal L}_{k0}^{GL}(T)/{\cal L}_{k}^{GL}(q,T)= 2\cos(2\phi/3)-1$, where $0 \le \phi \le \pi/2$ [see Eqs.\ (\ref{jsByjdGL})-(\ref{f2GL})]. The slow-experiment kinetic inductivity of the superfluid in the GL limit is 
\begin{equation}
{\cal L}_{k}^{GL}(q,T)=\mu_0\lambda_0^2(T)F_{s}^{GL}\Big(\frac{|j_s|}{j_{d}(T)}\Big),
\label{Lksq}
\end{equation}
where 
\begin{eqnarray}
F_{s}^{GL}(x) &=&\frac{1}{2\cos(2\phi/3)-1}
\label{FslowGL}
\end{eqnarray}
and $\phi = \sin^{-1}x$.  For small values of $x$, 
\begin{equation}
F_{s}^{GL}(x) = 1+ \frac{4}{9}x^2 + \frac{80}{243}x^4 + O(x^6),
\end{equation}
and $F_{s}^{GL}(x)$ diverges at $x = 1$, as noted in Ref.\ \onlinecite{Anlage89}.
 (See also the upper solid curve in Fig.\ \ref{Hfig}.)

\section{Superfluid kinetic inductivity in fast experiments (slow relaxation)\label{FastSec}}

We next consider fast experiments (or, equivalently, slow relaxation), in which the current density $j_s(t) = \bar j_s +j_{s1}(t)$ as a function of the time $t$ changes rapidly about its time average $\bar j_s$ on a time scale $\tau_{exp}$ much shorter than the relaxation time $\tau_s$.\cite{Anlage89,Tinkham96,Kopnin01}   In this case neither the order parameter $\Delta_q$ nor the $q$-dependent penetration depth $\lambda_q$ can follow the time dependence of the current, but instead they remain frozen to their values at  $q = \bar q$ given by $\bar j_s = j_{s \bar q} = -\bar A_s/\mu_0 \lambda_{\bar q}^2$, where $\bar A_s = m\bar v_s/e = \hbar \bar q/2e$.  From $j_s(t) =  -A_s(t)/\mu_0 \lambda_{\bar q}^2$, $A_s(t) = \bar A_s + A_{s1}(t),$  and $E(t) = -dA_s(t)/dt={\cal L}_{k}(\bar q,T)dj_{s}(t)/dt$, we find that the  kinetic inductivity of the superfluid in fast experiments is 
\begin{equation}
{\cal L}_{k}(\bar q,T)=\mu_0\lambda_{\bar q}^2(T)=\mu_0\lambda_0^2(T)F_{f}\Big(\frac{|\bar j_s|}{j_{d}(T)}\Big),
\label{LksFast}
\end{equation}
which can be evaluated numerically using Eq.\ (\ref{nsqsum})
as 
\begin{equation}
\frac{{\cal L}_{k}(\bar q,T)}{{\cal L}_{k0}(T)}=\frac{n_{s0}(T)}{n_{s\bar q}(T)}=\sum_{n=0}^\infty\frac{1}{1+u_{n0}^2}/\sum_{n=0}^\infty\frac{1}{1+u_{n\bar q}^2}.
\end{equation}
(See Fig.\ \ref{nsqtfig}.)
Here the fast-experiment function $F_f$ is simply ${\cal L}_{k}(\bar q,T)/\mu_0\lambda_0^2(T)$ but expressed as a function of the normalized current density $|\bar j_s|/j_{d}(T)$ rather than as a function of $\bar q$. When $\bar q \to 0$, ${\cal L}_{k}(\bar q,T)$ reduces to ${\cal L}_{k}(0,T)={\cal L}_{k0}(T)$ [Eq.\ (\ref{Lks0})].

\begin{figure}
\includegraphics[width=8cm]{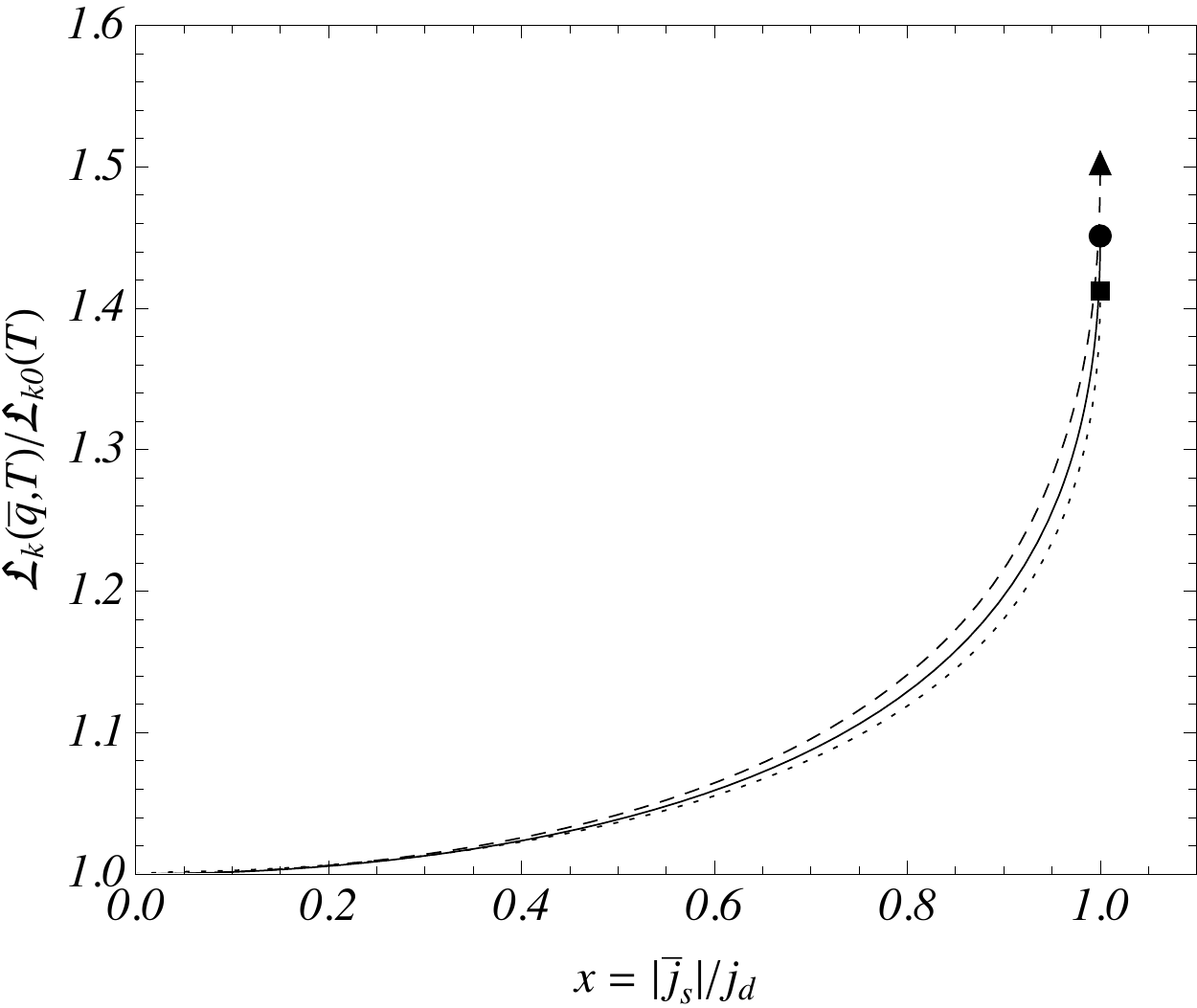}
\caption{%
${\cal L}_{k}(\bar q,T)/{\cal L}_{k0}(T)=F_f(|\bar j_s|/j_{d}(T))$ for fast experiments vs $x = |\bar j_s|/j_d(T)$ at $t = T/T_{c0} = 0$ (solid curve and filled  circle), $t =0.3$ (dotted curve and filled  square) and $t \to 1$ (dashed curve and filled  triangle). Note that ${\cal L}_{k0}(T) =\mu_0 \lambda_{0}^2(T)$. }
\label{Lksqffig}
\end{figure}

Shown in Fig.\ \ref{Lksqffig} is the typical dependence of  ${\cal L}_{k}(\bar q,T)/{\cal L}_{k0}(T)$ vs $|\bar j_s|/j_d(T)$.  This figure was obtained by (a)  evaluating $\tilde j_{\bar q}(T)$ and ${\cal L}_{k}(\bar q,T)/{\cal L}_{k0}(T)$ numerically for $t = T/T_{c0}$ = 0, 0.1, 0.2, 0.3, 0.4, 0.5, 0.6, 0.7, 0.8, and 0.9, and analytically in the GL limit $t\to 1$ [see Eqs.\ (\ref{jsByjdGL}) and  (\ref{f2GL})], and  (b) making a parametric plot of ${\cal L}_{k}(\bar q,T)/{\cal L}_{k0}(T)$ vs $|\bar j_s|/j_d = \tilde j_{\bar q}(T)/\tilde j_d(T)$.   As shown by the solid curve for $t = 0$, the dotted curve for $t = 0.3$, and the dashed curve for $t \to 1$, the behavior of ${\cal L}_{k}(\bar q,T)/{\cal L}_{k0}(T)$ vs $|\bar j_s|/j_d$ is not monotonic as the temperature changes, but the curves for all other temperatures (not shown) lie in a narrow band between the dotted and dashed curves.  As $|\bar j_s|/j_d \to 1$, all the curves approach their limiting values in the range 1.41 - 1.50 (solid symbols in Fig.\ \ref{Lksqffig}) with infinite slope.

The  curves shown in  Fig.\ \ref{Lksqffig} can be represented by $y_{fn}(x) = y_0-(y_0-1)(1-x^n)^{1/n}$ (not shown in Fig.\ \ref{Lksqffig}), which fits the calculated values of $y = {\cal L}_{k}(\bar q,T)/{\cal L}_{k0}(T)$ vs $x = |j_s| /j_d$, where $y_0$ is the value of ${\cal L}_{k}(\bar q,T)/{\cal L}_{k0}(T)$ at $x = 1$, for $0\le x < 0.97$ within 0.5\% for $(y_0,n,t)$ = (1.451, 2.48, 0), (1.448, 2.47, 0.1), (1.422, 2.45, 0.2), (1.412, 2.46, 0.3), (1.417, 2.50, 0.4), (1.432, 2.50, 0.5), (1.448, 2.50, 0.6), (1.463, 2.50, 0.7), (1.477, 2.50, 0.8), (1.490, 2.50, 0.9), and (1.500, 2.50, $t\to 1$).

To calculate ${\cal L}_{k}(\bar q,T)$ in the GL limit shown by the dashed curve in Fig.\  \ref{Lksqffig}, we used Eqs.\ (\ref{jsByjdGL})-(\ref{f2GL}) The kinetic inductivity of the superfluid for fast experiments in the GL limit is 
\begin{equation}
{\cal L}_{k}^{GL}(\bar q,T)=\mu_0\lambda_0^2(T)F_{f}^{GL}\Big(\frac{|\bar j_s|}{j_{d}}\Big),
\label{LksqfGL}
\end{equation}
where 
\begin{eqnarray}
F_{f}^{GL}(x) &=&\frac{1}{f^2(x)}=\frac{3}{1+2\cos(2\phi/3)} 
\label{FfastGL}
\end{eqnarray}
and $\phi = \sin^{-1}x$.
As noted in Ref.\ \onlinecite{Anlage89}, for small values of $x$, 
\begin{equation}
F_{f}^{GL}(x) = 1+ \frac{4}{27}x^2 + \frac{16}{243}x^4 + O(x^6),
\end{equation}
and $F_{f}^{GL}(x)$  approaches 3/2 with infinite slope as $x \to 1$.
 (See also the lower solid curve in Fig.\ \ref{Hfig}.)

\section{Kinetic impedivity of the superfluid ${\cal Z}_{ks}$\label{Complex}}

In the above sections we discussed the situations when $\tau_{exp}/\tau_s$ is large or small.  To describe in detail the transition between these two limits is well beyond the scope of this paper, because this topic involves nonequilibrium processes with numerous relaxation times.\cite{Kaplan76,Tinkham96,Kopnin01}  However, we present here an oversimplified procedure for approximating the transition between the two limits, although the actual transition is likely to be much more complicated.

For the moment we restrict our attention to the GL regime and employ a phenomenological model assuming that the time dependence of $f$  is determined by simplest version of the time-dependent GL (TDGL) equation,\cite{Tinkham96}
\begin{equation}
\tau_s df/dt = f-f^3-A_{s}'^2f,
\label{TDGL1}
\end{equation}
where  $A_{s}' = A_{s}/(\phi_0/2\pi\xi)$.  
An important caution here is that we are using this equation in the gapped state, even though a nonlinear TDGL equation has been rigorously justified only  in a gapless superconductor,\cite{Gorkov68} where near $T_c$ and at frequencies $\omega \tau_s \ll 1$, $\tau_s = \pi \hbar/8k_B(T_c-T)$.\cite{Tinkham96}

In the GL regime [$(T_c-T)\ll T_c$], the supercurrent density [Eq.\ (\ref{jsq})] becomes in dimensionless quantities
\begin{equation}
j_{s}' = -f^2 A_s',
\label{jsGL}
\end{equation} 
where $j_{s}' = j_{s}/(\phi_0/2\pi\mu_0\xi\lambda_0^2)$.
We now consider experiments in which the supercurrent density $j_s'(t)$ as a function of the time $t$ changes about its time average on a time scale $\tau_{exp}$ comparable with the relaxation time $\tau_s$. In particular, we consider the linear response of the superconducting strip to a time-dependent supercurrent density given by  $j_s'(t)=j_{s0}'+j_{s1}'e^{i\omega t}$, where $j_{s0}'$, the bias current current density, is fixed to be in the range $0 \le |j_{s0}'| < j_{d}'$, and $j_{s1}'$, the amplitude of the ac current density, obeys $j_{s1}' \ll j_{s0}'$. In this section we assume that the frequencies are sufficiently low that normal-fluid currents are not excited such that  the current is all supercurrent.  To analyze the linear response of the reduced order parameter to the ac current, we substitute $j_{s}' = j_{s0}' + j_{s1}' e^{i\omega t}$ and $f = f_0 + f_1 e^{i\omega t}$ ($|f_1| \ll f_0$) into Eq.\ (\ref{TDGL1}), where $f_0$ ($\sqrt{2/3} \le f_0 \le 1$) is the solution of Eq.\ (\ref{TDGL1}) in the time-independent case when  $j_{s}' = 
j_{s0}'$ .  We then linearize  Eq.\ (\ref{TDGL1}) by neglecting terms of order $\tilde j_{s1}^2$ and $f_1^2$.  The solution  is 
\begin{equation}
f_1 = -\frac{2j_{s0}'j_{s1}'}{f_0^3(6f_0^2-4+i\omega \tau_s)}.
\end{equation}
(Note that this result is obtained for sinusoidal variation of the supercurrent around a fixed value of $j_{s0}'$.  A different result would be obtained for sinusoidal variation of the gauge-invariant vector potential around a fixed value of  $A_{s0}'$.)

From  $E = -dA_s/dt$, Eq.\ (\ref{jsGL}), and $j_{s0}'^2 = f_0^4(1-f_0^2)$ we obtain  the electric field in the linear-response approximation:
\begin{eqnarray}
E &=& \mu_0 \lambda_0^2 \Big(\frac{2f_0^2+i\omega \tau_s}{f_0^2(6f_0^2-4+i\omega\tau_s)}\Big)\frac{dj_{s1}}{dt}\label{Ewithf0}\\
&=&{\cal Z}_{ks} j_{s1} = ({\cal R}_{ks} +i{\cal X}_{ks})j_{s1}\\
&=& ({\cal R}_{ks} +i\omega{\cal L}_{k})j_{s1}.
\end{eqnarray}
The coefficient of $dj_{s1}/dt$ on the right-hand side of Eq.\ (\ref{Ewithf0}) reduces to the slow-experiment inductivity ${\cal L}_{k}^{GL}(q,T)$ [Eq.\ (\ref{Lksq})] in the limit $\omega \tau_s \to 0$ and to the fast-experiment inductivity ${\cal L}_{k}^{GL}(\bar q,T)$ [Eq.\ (\ref{LksqfGL})] in the limit $\omega \tau_s \to \infty.$

The  complex kinetic impedivity (specific impedance or complex resistivity) of the superfluid ${\cal Z}_{ks}$, here evaluated in the GL regime, can be conveniently expressed in terms of $F_s^{GL}$ [Eq.\ (\ref{FslowGL})] and $F_f^{GL}$ [Eq.\ (\ref{FfastGL})] as
\begin{equation}
{\cal Z}_{ks}=i\omega\mu_0 \lambda_0^2 \Big(\frac{F_s^{GL}+F_f^{GL}i\omega \tau_{eff}}{1+i\omega \tau_{eff}}\Big),
\label{calZ}
\end{equation}
where
\begin{equation}
\tau_{eff}= F_s^{GL}\tau_s/2.
\label{taueff}
\end{equation}
(For a long strip of length $\ell$, width $W$, and thickness $d$, the complex kinetic impedance is $Z_{ks} = {\cal Z}_{ks} \ell/Wd$.)

The real part ${\cal R}_{ks}$ of the superfluid kinetic impedivity  is the frequency-dependent resistivity  of the superfluid due to order-parameter relaxation.  Using the  the parametric relation $x = |j_{s0}|/j_d = \sin\phi$ as above, we obtain
\begin{eqnarray}
{\cal R}_{ks}&=& \frac{\mu_0\lambda_0^2}{\tau_s}G\Big(\frac{|j_{s0}|}{j_d},\omega \tau_s \Big),
\label{calR}\\
G(x,\omega \tau_s)&=& \frac{\beta(\phi) (\omega \tau_s)^2}{\alpha^2(\phi) + (\omega \tau_s)^2},
\label{G}\\
\alpha(\phi)&=&4\cos(2\phi/3)-2,\\
\beta(\phi)&=& \frac{16\sin^2(\phi/3)}{1+2\cos(2\phi/3)}.
\end{eqnarray}
\begin{figure}
\includegraphics[width=8cm]{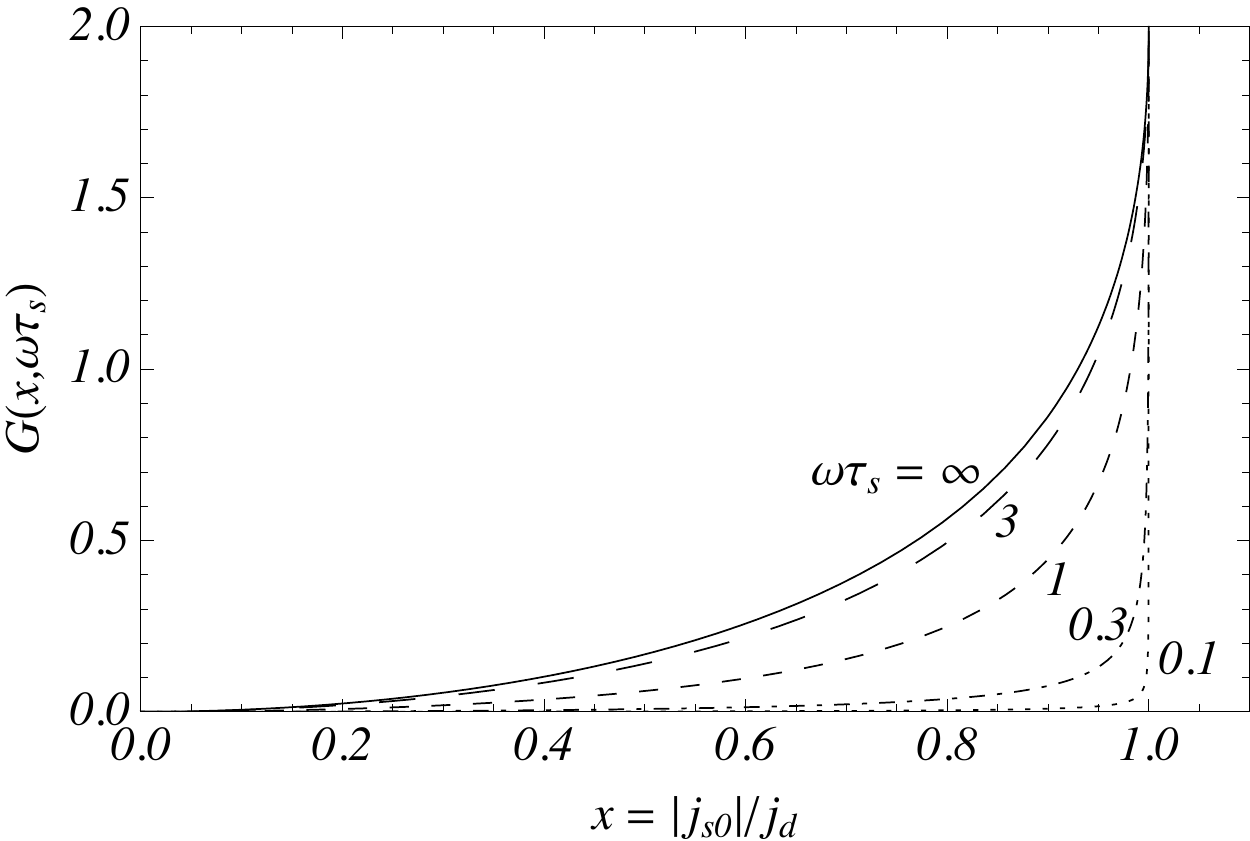}
\caption{%
$G(x,\omega\tau_s)$, which describes the alternating-current resistivity  of the superfluid due to order-parameter relaxation [Eqs.\ (\ref{calR}) and (\ref{G})], vs $x = |j_{s0}|/j_d$ for  $\omega \tau_s = 0.1$ (dotted), $\omega \tau_s = 0.3$ (dot-dashed), $\omega \tau_s = 1$ (dashed),   $\omega \tau_s = 3$ (long dash), and $\omega \tau_s = \infty$.  $G(1,\omega\tau_s)=2.$ }
\label{Gfig}
\end{figure}
$G(x,\omega \tau_s)$ is shown in Fig.\ \ref{Gfig} as a function of $x$ for various values of $\omega\tau_s$.  In limiting cases, we have
\begin{eqnarray}
\!\!\!\!\!\!G(x,\omega \tau_s)&=& \frac{16}{27}x^2+ \frac{64}{243}x^4 +O(x^6),\;\omega \tau_s \gg 1,\\
&=&\Big[\frac{4}{27}x^2+ \frac{16}{81}x^4 +O(x^6)\Big](\omega \tau_s)^2,\nonumber\\
&&\;\;\;\;\;\;\;\;\;\;\;\;\;\;\;\;\;\;\;\;\;\;\;\;\;\;\;\;\;\;\;\;\;\;\;\;\;\;\;\omega \tau_s \ll 1,
\end{eqnarray}
and $G(x,\omega \tau_s)$  approaches 2 with infinite slope as $x \to 1$.  

Although the superconducting strip has zero dc electrical resistivity, under ac conditions order-parameter relaxation  contributes to dissipation of energy in a manner similar to the way it contributes to flux-flow dissipation.\cite{Tinkham96,Tinkham64,Schmid66,Gorkov75,Dorsey92}  The time-averaged rate of energy dissipation per unit volume via order-parameter relaxation is $(1/2){\cal R}_{ks} j_{s1}^2$, which also can be calculated using the dissipation function discussed in Refs.\ \onlinecite{Schmid66} and \onlinecite{Gorkov75}.

\begin{figure}
\includegraphics[width=8cm]{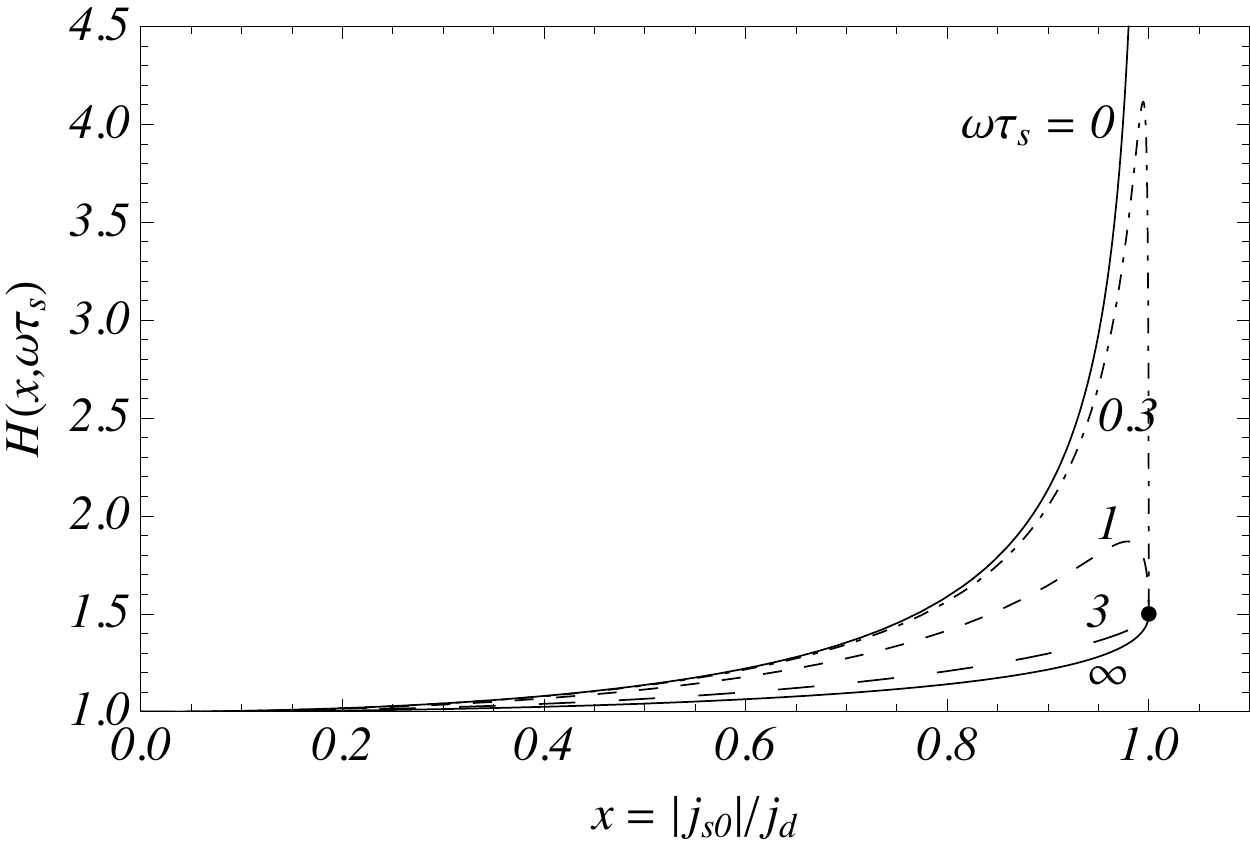}
\caption{%
$H(x,\omega\tau_s)$, which describes the superfluid's kinetic inductivity [Eqs.\ (\ref{LksH}) and (\ref{H})], vs $x = |j_{s0}|/j_d$ for  $\omega \tau_s = 0$ [upper solid curve, for which $H(x,0)=F_s^{GL}(x)$, Eq.\ (\ref{FslowGL})],  $\omega \tau_s = 0.3$ (dot-dashed), $\omega \tau_s = 1$ (dashed),   $\omega \tau_s = 3$ (long dash), and $\omega \tau_s = \infty$ [lower solid curve, for which $H(x,0)=F_f^{GL}(x)$, Eq.\ (\ref{FfastGL})]. For nonzero $\omega \tau_s$, $H(1,\omega\tau_s)=1.5$ (black point). }
\label{Hfig}
\end{figure}

The superfluid's kinetic reactivity is ${\cal X}_{ks}=\omega {\cal L}_{k}$, where the superfluid's kinetic inductivity ${\cal L}_{k}$ is
\begin{eqnarray}
{\cal L}_{k}&=& \!\!\mu_0\lambda_0^2H\Big(\frac{|j_{s0}|}{j_d},\omega \tau_s \Big),
\label{LksH}\\
\!\!\!\!\!\!\!\!\!H(x,\omega\tau_s)\!\!\!&=& \!\!\!\frac{1}{1\!+\!2\cos(2\phi/3)}\Big[3\!+\!\frac{16\alpha(\phi)\sin^2(\phi/3)}{\alpha^2(\phi) + (\omega \tau_s)^2}\Big].
\label{H}
\end{eqnarray}
Shown in Fig.\ \ref{Hfig} are plots of $H(x,\omega\tau_s)$ vs $x$ for various values of  $\omega \tau_s$.  As expected,  $H(x,\omega\tau_s)$ approaches the slow-experiment result $F_s^{GL}(x)$ when $\omega \tau_s \to 0$ and  the fast-experiment result $F_f^{GL}(x)$  as $\omega \tau_s \to \infty $.

In the above we have used the relatively simple formalism of the time-dependent GL theory, bearing in mind that there are questions of whether this theory can legitimately be used for gapped superconductors and how $\tau_s$ can be determined.  A reasonable starting point for an approximate phenomenological theory of  the complex kinetic impedivity of the superfluid ${\cal Z}_{ks}$ at lower temperatures  outside the GL regime would be to replace the quantities $F_s^{GL}$ and $F_f^{GL}$ in  Eqs.\ (\ref{calZ}) and (\ref{taueff}) by the more general expressions $F_s$ and $F_f$ given in Eqs.\ (\ref{LksSlow}) and (\ref{LksFast}).  Note from Figs.\ \ref{invLksqfig} and \ref{Lksqffig} that $F_s={\cal L}_{k}(q,T)/\mu_0\lambda_0^2(T)$ and $F_f={\cal L}_{k}(\bar q,T)/\mu_0\lambda_0^2(T)$ as functions of $|j_s|/j_d$ and $|\bar j_s|/j_d$ do not differ greatly from their GL counterparts, $F_s^{GL}$ and $F_f^{GL}$. Although the theory we have presented here is not rigorous, our results suggest that when $\omega \tau_s \gg 1$ but when  $\omega$ is well below the superconducting gap frequency $2\Delta_q(T)/\hbar$, order-parameter relaxation gives rise to a current- and frequency-dependent contribution to the ac resistivity separate from that due to the normal fluid (thermally excited quasiparticles).  
However,  as discussed above, remaining  unknown is how to calculate order-parameter relaxation  and how to determine the relaxation times that should replace $\tau_s$ in a more complete theory.\cite{Kopnin01}

\section{Kinetic impedance including the normal-fluid response \label{Normal}}

We examine next the current dependence of the dissipation arising from the flow of thermally excited quasiparticles at frequencies $\omega$ well below the superconducting gap frequency $2\Delta_q(T)/\hbar$. This is the frequency regime where a two-fluid approach is generally applicable.\cite{Tinkham96}   However, the two-fluid terminology   needs to be used with caution, because, as explained below, coherence-factor effects  can produce dissipation greater than in the normal state.\cite{Marsiglio91} Although the quasiparticles have their own kinetic inductivity,\cite{Meservey69} their reactive contribution to the  total kinetic impedivity is negligible at the frequencies of interest here.  The quasiparticles' only significant contribution to the ac normal-fluid current density is therefore
\begin{equation}
j_{n1} = \sigma_{1}E,
\end{equation}
where $\sigma_1$ corresponds to the real part of the complex conductivity $\sigma = \sigma_1 -i \sigma_2$, the linear response function connecting $\bm j$ and $\bm E$ calculated by Mattis and Bardeen.\cite{Mattis58} 

We begin by rexpressing $\sigma_1$ for the BCS case as\cite{Clem66} 
\begin{eqnarray}
\sigma_1 \!\!&=&\!\! \frac{\sigma_n}{\omega}\int_{-\infty}^{\infty}d\omega'[f(\omega'-\frac{\omega}{2})-f(\omega'+\frac{\omega}{2}]
\nonumber \\
&\times&\!\!\![n_1(\omega'\!-\!\frac{\omega}{2})n_1(\omega'\!+\!\frac{\omega}{2})+p_1(\omega'\!-\!\frac{\omega}{2})p_1(\omega'\!+\!\frac{\omega}{2})],\qquad
\label{sigma1}
\end{eqnarray}
where $\omega$ is expressed in energy units ($\hbar = 1$) and $f(\omega) = 1/(1+e^{\beta \omega})$ is the Fermi function.  For the BCS case,
\begin{eqnarray}
n_1(\omega)&=&{\rm Re} \frac{\omega}{\sqrt{\omega^2-\Delta^2}}, \label{n1}\\
p_1(\omega)&=&{\rm Re} \frac{\Delta}{\sqrt{\omega^2-\Delta^2}}, \label{p1}
\end{eqnarray}
$n_1(\omega)=p_1(\omega) = 0$ for $|\omega|< \Delta(T)$, and the signs of the square roots are chosen such that $n_1(\omega)$ is an even function and $p_1(\omega)$ an odd function of $\omega$.
At $T = 0$, the Fermi functions freeze out all contributions to $\sigma_1$ for $\omega < 2\Delta(0)$, such that $\sigma_1$ is nonvanishing only for  $\omega > 2\Delta(0)$, when the integral of Eq.\ (\ref{sigma1}) can be expressed in terms of complete elliptic integrals.\cite{Mattis58,Tinkham96}  
For $T > 0$ and frequencies obeying $\omega \ll 2\Delta(0)$, $\sigma_1/\sigma_n$ plotted as a function of temperature is found theoretically in dirty superconductors\cite{Marsiglio91} to be very small at low temperatures, rising to a maximum  at which $\sigma_1/\sigma_n > 1$ (for example,\cite{Marsiglio91} $\sigma_{1max}/\sigma_n$ = 2.17 at $t = 0.864$ when $\omega/\Delta(0) = 0.02$)  and returning to 1 at $t = T/T_{c0}=1$. 
A similar temperature dependence has been seen experimentally in several materials.\cite{Holczer91a,Holczer91b,Klein94,Jin03}

Because the coherence-factor terms $n_1$ and $p_1$  in Eq.\ (\ref{sigma1}) yield a logarithmic divergence for $\omega = 0$, $\sigma_1$ can be evaluated approximately at low temperatures [$\Delta(T)/k_BT \gtrsim 2$] and low frequencies [$\omega \ll 2\Delta(0)$] by introducing a cutoff energy  $\epsilon \sim \omega$.  The leading term in the result is 
\begin{equation}
\sigma_1 \approx 2 \sigma_n \Big[\frac{\Delta(T)}{k_B T}\Big]\exp\Big[-\frac{\Delta(T)}{k_B T}\Big]\ln\Big(\frac{k_B T}{\epsilon}\Big).
\label{sigma1approx}
\end{equation}

The expressions for $\sigma_1$ derived by Nam\cite{Nam67a,Nam67b,Nam70} for strong-coupling superconductors also can be put into the form of Eq.\ (\ref{sigma1}), except that $\Delta$ in  Eqs.\ (\ref{n1}) and (\ref{p1}) must then be replaced by the complex gap function $\Delta(\omega)$, which contains additional phonon-related $\omega$ dependence due to the electron-phonon interaction.\cite{Eliashberg60} 

For the pair-breaking theory of superconductors with paramagnetic impurities\cite{Skalski64,Ambegaokar65,Nam67a,Nam67b,Nam70} or, as in the case of interest here, current-carrying thin films,\cite{Maki63a,Maki63b,Maki64,Maki65,Maki67,Maki69,Sridhar86} Eqs.\ (\ref{n1}) and (\ref{p1}) are replaced by
\begin{eqnarray}
n_1(\omega)&=&{\rm Re} \frac{u}{\sqrt{u^2-1}}, \label{n1u}\\
p_1(\omega)&=&{\rm Re} \frac{1}{\sqrt{u^2-1}}, \label{p1u}
\end{eqnarray}
where $u$ is given by Eq.\ (\ref{uofomega}), $n_1(\omega)=p_1(\omega) = 0$ for $|\omega|< \omega_g$ with\cite{Skalski64} $\omega_g/\Delta_q = (1-\zeta^{2/3})^{3/2}$, and the signs of the square roots are chosen such that $n_1(\omega)$ is an even function and $p_1(\omega)$ an odd function of $\omega$.  (See, for example, Fig.\ 4 in Ref.\ \onlinecite{Skalski64} or Fig.\ 6 in Ref.\ \onlinecite{Ambegaokar65}.)  Numerical evaluation of  Eq.\ (\ref{sigma1}) using Eqs.\ (\ref{n1u}) and (\ref{p1u}) in the limit as $\zeta \to 0$ yield $\sigma_1(\omega)$ values in agreement with the dirty-limit Mattis-Bardeen results.\cite{Mattis58,Tinkham96} 

\begin{figure}
\includegraphics[width=8cm]{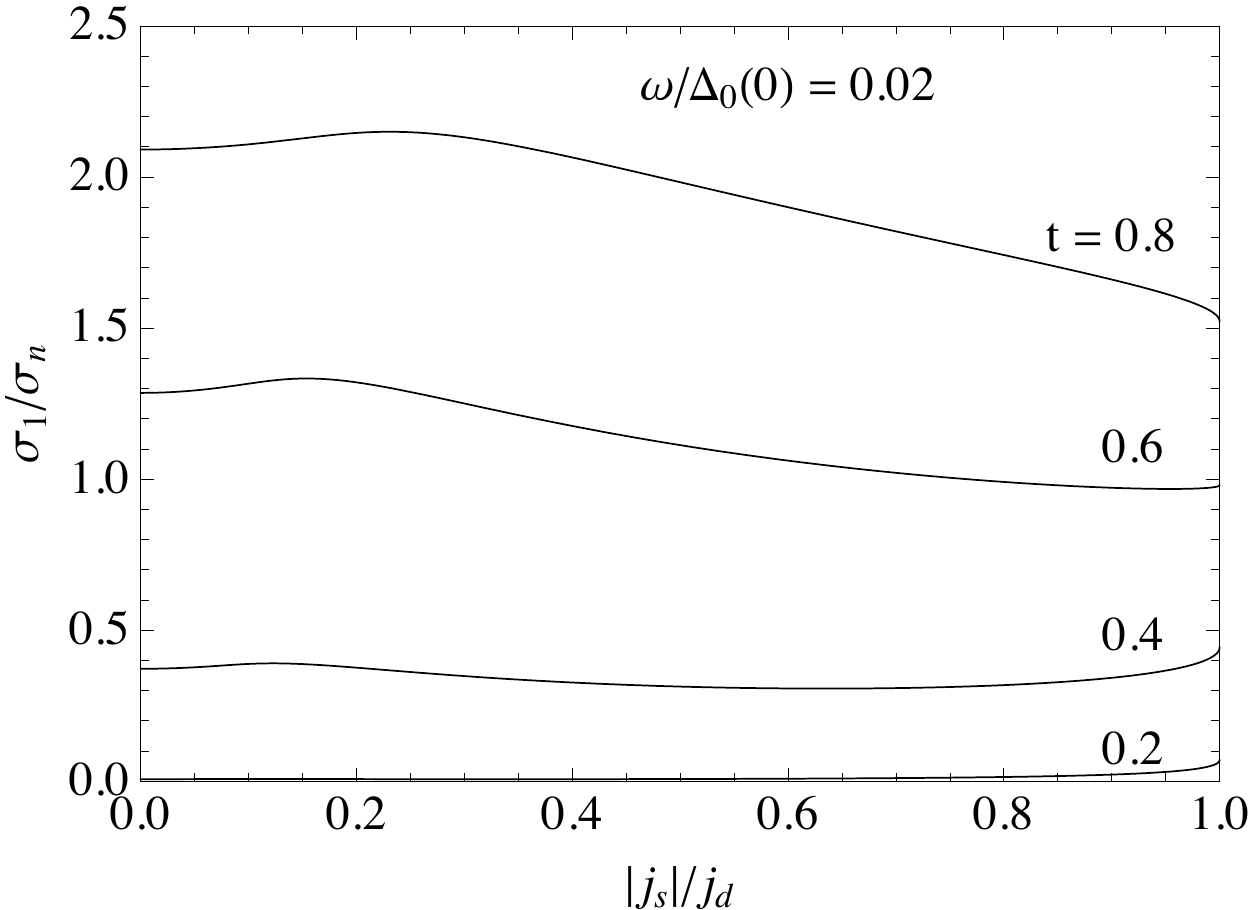}
\caption{%
$\sigma_1/\sigma_n$ vs $|j_s|/j_d$ calculated from Eqs.\ (\ref{sigma1}), (\ref{n1u}), and (\ref{p1u}) for $\omega/\Delta_0(0) = 0.02$ and $t = T/T_{c0} = $ 0.2, 0.4, 0.6, and 0.8.}
\label{sigma1fig}
\end{figure}

Figure \ref{sigma1fig} shows $\sigma_1/\sigma_n$ vs $|j_s|/j_d$ for $\omega/\Delta_0(0) = 0.02$ and several values of  $t = T/T_{c0}$.  $\sigma_1/\sigma_n$ was first calculated from  Eq.\ (\ref{sigma1}) as a function of $q/q_m(0)$ up to $q_d(T)/q_m(0)$, accounting for the $q$ and temperature dependence of $\zeta = (q^2/2q_m^2)\Delta_0(0)/\Delta_q(T).$  (See Fig.\ \ref{Deltaqtfig}.)  Figure \ref{sigma1fig} was then constructed as a parametric plot using $|j_s|/j_d$ vs $q/q_m(0)$  up to $q_d(T)/q_m(0)$ obtained as in Fig.\ \ref{tildejqtfig}. 
The main features of the dependence of $\sigma_1/\sigma_n$ can be understood from Eq.\ (\ref{sigma1approx}), but where $\Delta(T)$ is replaced by $\Delta_q(T)$ (see Fig.\ \ref{Deltaqtfig}) and the cutoff $\epsilon$ is replaced by the larger of $\epsilon_\omega \sim \omega$ or the peak width in $n_1(\omega)$, $\epsilon_\zeta \sim \Delta_q -\omega_g \approx (3/2)\Delta_q\zeta^{2/3}$.  The exponential term in Eq.\ (\ref{sigma1approx}) tends to make $\sigma_1$ increase as $q$ and $|j_s|$ (see  Figs.\ \ref{tildejqtfig} and \ref{jq0fig}) increase, as seen in Fig.\ \ref{sigma1fig} for small values of $|j_s|/j_d$, where $\epsilon_\omega > \epsilon_\zeta$ and the $\ln(k_BT/\epsilon_\omega)$ term is a constant.  Maxima occur when $\epsilon_\omega \approx \epsilon_\zeta$.  To the right of the maxima, $\epsilon_\zeta > \epsilon_\omega$, and the term $\ln(k_BT/\epsilon_\zeta)$, which is a decreasing function of $q$ and $|j_s|$, plays a stronger role.

The current density carried by the normal fluid, accounting for the effects of the coherence factor, is 
\begin{equation}
j_{n1} = \sigma_1 E.
\end{equation}
We also can express the normal-fluid response in terms of the normal-fluid's kinetic  impedivity ${\cal Z}_{kn}= 1/\sigma_1$.  Because the total ac current density carried by the strip is $j_1 = j_{s1} + j_{n1}$, and $E = {\cal Z}_{k}j_{s1}$, the overall kinetic  impedivity ${\cal Z}_k = {\cal R}_k +i{\cal X}_k = {\cal R}_k +i\omega {\cal L}_k$ of the strip, including the normal-fluid response, is the impedances-in-parallel combination,
\begin{equation}
{\cal Z}_k = ({\cal Z}_{ks}^{-1}+{\cal Z}_{kn}^{-1})^{-1}.  
\end{equation} 
Note that as $T\to T_{cq}$, $|{\cal Z}_{ks}|$ diverges,  ${\cal Z}_{kn} \to 1/\sigma_n$, and ${\cal Z}_k$ approaches the normal-state resistivity. 

In the limit as $j_{s0} \to 0$, our results reduce to the well-known two-fluid description\cite{Tinkham96}  when the impedivity can be expressed in terms of the  complex conductivity $\sigma = \sigma_1 - i\sigma_2$, the linear response function connecting $\bm j$ and $\bm E$ calculated by Mattis and Bardeen.\cite{Mattis58}  As $\bm j_{s0} \to 0$, ${\cal R}_{ks} \to 0$, ${\cal X}_{ks} \to \omega {\cal L}_{ks} = \mu_0\omega \lambda_0^2 = 1/\sigma_2$, $\sigma_{nf} \to \sigma_1$, ${\cal Z}_{ks} \to i/\sigma_2,$ ${\cal Z}_{kn} \to 1/\sigma_1$, and ${\cal Z}_k \to (\sigma_1 -i\sigma_2)^{-1} = \sigma^{-1}$.  

At temperatures above $T_{cq}$, the normal-state impedance is $Z = R_n + i \omega (L_k+L_m)$, where $L_m$ is the geometric inductance  associated with stored magnetic energy and, for a long strip of conduction-electron density $n_c$, total length $\ell$, width $W$, and thickness $d$,  $L_k= (m/n_ce^2)(\ell/Wd)$ is the normal-state kinetic inductance.\cite{Meservey69} The impedance is usually dominated by the normal-state resistance $R_n =\rho_n \ell/Wd$ except at very high frequencies.\cite{Meservey69}

\section{Discussion\label{discussion}}

In this paper we have presented fundamental theoretical calculations of the kinetic impedance of thin and narrow  impure superconducting films for all temperatures and for all currents up to the depairing current.  Our results should be applicable to ongoing experimental studies of small-scale superconducting devices in which the kinetic inductance plays an important role. Our calculations have shown examples of how the kinetic inductance and the normal-fluid dissipation depend upon the dc applied current.  
However, experiments examining the in-phase and out-of-phase third and higher harmonics might provide a more sensitive means of revealing the influence of the nonlinearities implied by these current dependencies. \cite{AnlageFoot}

Our results in the GL regime for the bias-current dependence of the kinetic inductance are in agreement with those of Anlage {\it et al.}\cite{Anlage89} and Annunziata {\it et al.}\cite{Annunziata10b} for the slow-experiment case and with the result of Anlage {\it et al.}\cite{Anlage89} for the fast-experiment case.   However, Annunziata {\it et al.},\cite{Annunziata10b} in examining the case of $T = 0$ and noting correctly that $\lambda_0^2(0)$  is inversely proportional to $\Delta_0(0)$ (in our notation) [see Eq.\ (\ref{lambda0T&sigman})], assumed that $\lambda_q^2(0)$  is inversely proportional to $\Delta_q(0)$.  However, this assumption is incorrect, as can be seen from Eqs.\ (\ref{deltaless}), (\ref{nsq01}), and (\ref{nsqbyns0}).  As a consequence, their prediction for the current dependence of the kinetic inductance does not agree with our results for either slow or fast experiments. 

In thin and narrow strips with sharp corners, current crowding leads to suppression of the order parameter in the immediate vicinity of sharp inner corners, and this can cause the critical current in such devices to be considerably lower than the depairing value.\cite{Clem11,Clem12,Hortensius12,Henrich12,Akhlaghi12,Berdiyorov12} By optimally rounding the inner corners, one should be able to raise the critical current to values close to the depairing value.\cite{Clem11,Hortensius12,Akhlaghi12}

As discussed in Sec.\ \ref{jsSec}, our calculation of the critical depairing current density $j_d$ has been carried out within a mean-field approach disregarding fluctuations. 
However, the experimental critical current density $j_c$ could be somewhat smaller than $j_d$ as a result of thermal or quantum fluctuations, which can initiate phase slips in 1D wires\cite{Langer67,McCumber70} or vortex nucleation at the edges of  strips\cite{Tafuri06,Bulaevskii11,Bulaevskii12,Clem11} when an energy barrier is overcome or suppressed to zero.   The fluctuation-limited $j_c$ therefore may prevent the observation of both the predicted divergence of the slow-experiment kinetic inductivity at $j_d$ and the approach to the maximum values of the fast-experiment kinetic inductivity (shown by the filled symbols in Fig.\ \ref{Lksqffig}).  
In fact, previous experimental observations of a relatively small kinetic inductance  rising to a peak and then rapidly dropping to smaller values with increasing applied current\cite{Enpuku95,Johnson97,Johnson98,Annunziata10b} are explainable in terms of the growth of high-resistance normal regions as the current density rises above $j_c$.

As discussed in Secs.\ \ref{SlowSec}-\ref{Complex}, relaxation of the superfluid, characterized approximately here by the relaxation time $\tau_s$, plays an important role in determining both the real and imaginary parts of the kinetic impedance.  To examine the  physics of relaxation dynamics is beyond the scope of this paper, and for  further discussion we refer the reader to Refs.\ \onlinecite{Tinkham96} and \onlinecite{Kopnin01} and references therein.   Nevertheless, Figs.\ \ref{Gfig} and \ref{Hfig} suggest means by which $\tau_s$ could be estimated from experimental determinations of the superfluid's ac resistivity and inductivity.  At the very least, such experiments should be able to reveal whether they are in the slow- or fast-experiment limit, and experiments carried out at different frequencies and temperatures might be able to show the transition between these two limits.

In this paper we have used the concept of the $q$-dependent penetration depth, which increases as the current density increases.  This occurs because an increase of the current density causes a decrease in the magnitude of the superconducting order parameter.  This concept is the basis of the effective field-dependent penetration depth in type-II superconductors $\lambda(B,T)$, which diverges as $B \to B_{c2}(T)$. This quantity has been introduced to understand such phenomena as small-angle neutron-scattering form factors,\cite{Clem75a} magnetic coupling of vortex lattices in dc superconducting transformers,\cite{Clem75b,Ekin75} magnetization curves in low-pinning superconductors,\cite{Hao91}  elastic properties of the vortex lattice,\cite{Brandt86}  and $\mu$SR measurements in the mixed state.\cite{Yaouanc97,Sonier07}

Here we have used the wavevector $q = 2\pi A_s/\phi_0$ to discuss the nonlinear response of the supercurrent to the gauge-invariant vector potential $A_s$ in the local limit.  It is important not to confuse our $q$ with the $q$ used by Tinkham\cite{Tinkham96} in his discussion of the nonlocal electrodynamics of superconductors, where it arises from Fourier transforming the convolution integral relating the linear response of the current density $\bm j(\bm r)$ to the vector potential $\bm A(\bm r')$, as calculated by Mattis and Bardeen.\cite{Mattis58}

\section*{ACKNOWLEDGMENTS} 
We thank K. K. Berggren, Y. Mawatari, D. Prober, D. Santavicca, and S. M. Anlage for stimulating suggestions and comments.
This research was supported by the U.S.\ Department of
Energy, Office of Basic Energy Science, Division of Materials
Sciences and Engineering and was performed at
the Ames Laboratory, which is operated for the U.S.\ Department
of Energy by Iowa State University under Contract No.
DE-AC02-07CH11358.  

\appendix

\section{$u_{nq}$\label{unq}}

The function $u_{n q}(\eta,\epsilon,\zeta)$ is the solution of the quartic equation 
\begin{equation}
\epsilon^2 u^4-2\eta \epsilon u^3 +(\eta^2+\epsilon^2-\epsilon^2\zeta^2) u^2 -2\eta \epsilon u -\eta^2 = 0,
\end{equation}
obtained from Eq.\ (\ref{unqomegaq}).   The desired solution of the quartic equation is 
\begin{equation}
u_{n q}=(f+g+\eta/\epsilon)/2,
\label{unqsoln}
\end{equation}
where
\begin{eqnarray}
a&=&-2[\epsilon^6(\zeta^2-1)^3-3\epsilon^4(1+16\zeta^2+\zeta^4)\eta^2\nonumber\\
&+&3\epsilon^2(\zeta^2-1)\eta^4-\eta^6],\\
b&=&\sqrt{-4c^6+a^2},\\
c&=&\epsilon^2(\zeta^2-1)-\eta^2,\\
d&=&(a+b)^{1/3},\\
e&=&\frac{2^{2/3}c^2+d^2}{ 2^{1/3}3\epsilon^2d},\\
f&=&\sqrt{\frac{2\epsilon^2(\zeta^2-1)+\eta^2}{3\epsilon^2} +e},\\
g&=&\sqrt{\frac{2[2\epsilon^2(\zeta^2-1)+\eta^2]}{3\epsilon^2}-e+\frac{2(\zeta^2+1)\eta}{\epsilon f}}.
\end{eqnarray}

Similar solutions of the quartic equation that arises in the closely related problem of the density of states in the Abrikosov-Gor'kov theory\cite{AG61} were obtained in Refs.\ \onlinecite{Nam67b} and \onlinecite{Srivistava08}.

\section{Zero-current limit\label{qto0}}

In the limit of zero current, $q\to 0$ ($v_s \to 0$), $u_{n 0}=\eta/\epsilon$, and we obtain from Eq.\ (\ref{selfcons}) after defining $t = T/T_{c0}$ and $\delta_0(T) = \Delta_0(T)/2\pi k_B T_{c0}$, 
\begin{eqnarray}
 \ln\frac{1}{t}= 
 \sum_{n=0}^{\infty}\Big(\frac{1}{n+1/2}-\frac{1}{\sqrt{(n+1/2)^2+\delta_0^2(T)/t^2}}\Big), \label{Delta_0}
\end{eqnarray}
which yields the temperature dependence of $\Delta_0(T)$  for all temperatures between 0 and $T_{c0}$, the transition temperature for $q = 0$.  An analysis of this equation as $t \to 0$ reveals that $4\pi\delta_0(0) = 2\Delta_0(0)/k_B T_{c0} = 2\pi e^{-\gamma}=3.528$, which is consistent with the fact that the above form of the Usadel theory coincides with the weak-coupling BCS theory for an s-wave isotropic gap on a spherical Fermi surface.  Values of $\Delta_0(T)/\Delta_0(0)$ vs $t = T/T_{c0}$, which reproduce well-known results,\cite{BCS57,Muehlschlegel59} can be  obtained by numerically carrying out the sum in Eq.\ (\ref{Delta_0}), and it  can be shown analytically that   $[\Delta_0(T)/\Delta_0(0)]^2 \to [8e^{2\gamma}/7\zeta(3)](1-t) = 3.016(1-t)$ as $t \to 1$. 

\begin{figure}
\includegraphics[width=8cm]{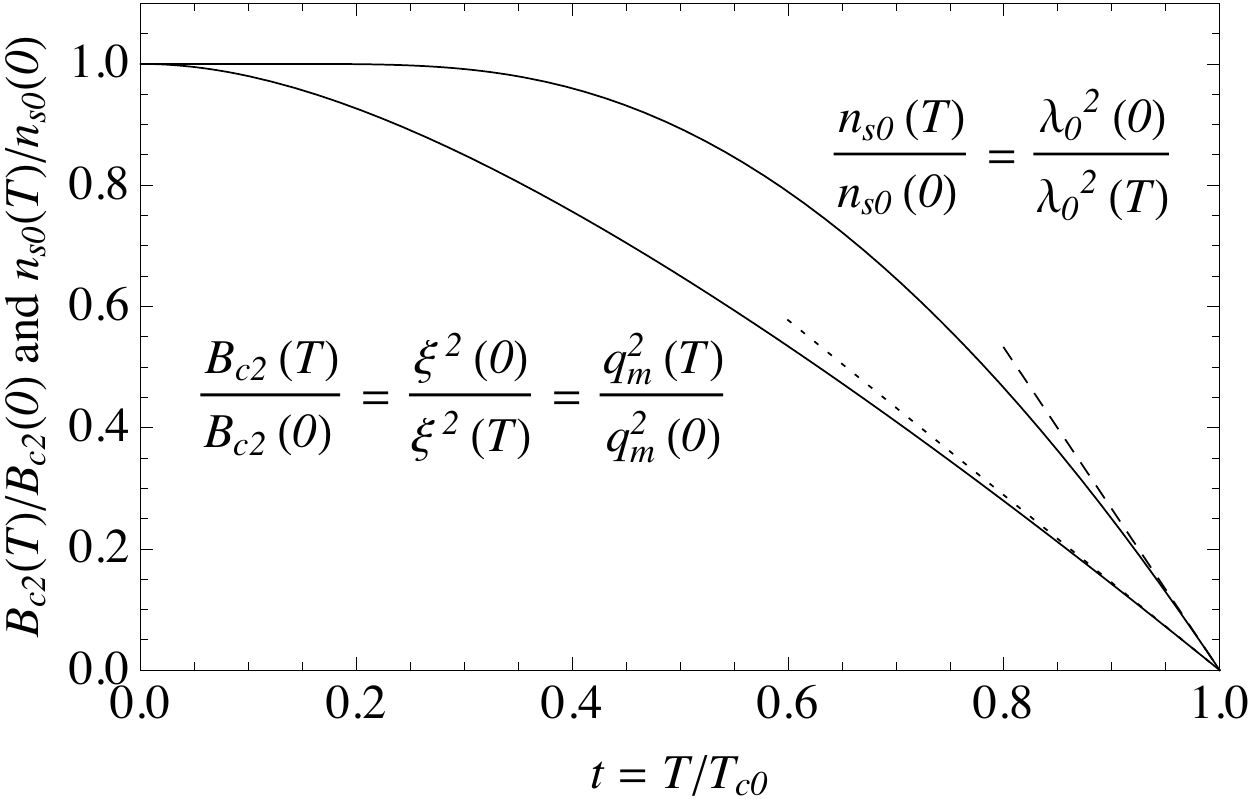}
\caption{%
Upper curve: reduced superfluid density $n_{s0}(T)/n_{s0}(0)=[\lambda_0(0)/\lambda_0(T)]^2$  vs $t=T/T_{c0}$, obtained from Eq.\ (\ref{ns0T}). The dashed line, $2.660 (1-t)$, shows the slope as $t \to 1$.  Lower curve:  reduced upper critical field $B_{c2}(T)/B_{c2}(0)=[\xi(0)/\xi(T)]^2$  vs $t=T/T_{c0}$, obtained from Eqs.\ (\ref{Bc21}) and (\ref{Bc22}). The dotted line, $1.444 (1-t)$, shows the slope as $t \to 1$. }
\label{nsofig}
\end{figure}

In the zero-$q$  limit, the  sum in Eq.\ (\ref{nsqsum}) can be evaluated analytically as shown in Eq.\ (\ref{ns0T}), such that, as previously discussed in Refs.\  \onlinecite{AGD65} and \onlinecite{Tinkham96},
\begin{equation}
\frac{n_{s0}(T)}{n_{s0}(0)}=\frac{\lambda_0^2(0)}{\lambda_0^2(T)} = \frac{\Delta_0(T)}{\Delta_0(0)}
\tanh\Big[\frac{\Delta_0(T)}{2k_B T}\Big].
\label{ns0TByns00}
\end{equation}
From Eq.\ (\ref{Lks0}), $\xi_0 = \hbar v_F/\pi \Delta_0(0)$, and 
the normal-state conductivity $\sigma_n =  2e^2N(0)D$, we obtain
\begin{equation}
\frac{1}{\mu_0\lambda_0^2(T)}=\frac{n_{s0}(T)e^2}{m} = \frac{\pi\sigma_n\Delta_0(T)}{\hbar}
\tanh\Big[\frac{\Delta_0(T)}{2k_B T}\Big].
\label{lambda0T&sigman}
\end{equation} 
Figure \ref{nsofig} exhibits the temperature dependence of $n_{s0}(T)$.  As $t \to 1$, $n_{s0}(T)/n_{s0}(0) \to [4\pi e^\gamma/7\zeta(3)](1-t) = 2.660 (1-t)$.  

At $T = 0$, we can write 
\begin{equation}
\frac{1}{\mu_0\lambda_0^2(0)}=\frac{2}{3}N(0)e^2v_F^2\Big(\frac{\ell}{\xi_0}\Big),
\label{lambda002}
\end{equation} 
but because the zero-temperature London penetration depth $\lambda_L(0)$ can be expressed as\cite{BCS57}
\begin{equation}
\frac{1}{\mu_0\lambda_L^2(0)}=\frac{2}{3}N(0)e^2v_F^2,
\label{lambdaL02}
\end{equation}
we see that $\lambda_0(0)=\lambda_L(0)(\xi_0/\ell)^{1/2}$.  (Recall that $\ell \ll \xi_0$ in the dirty limit under consideration here.)

\section{Zero-gap limit \label{Deltato0}}

\begin{figure}
\includegraphics[width=8cm]{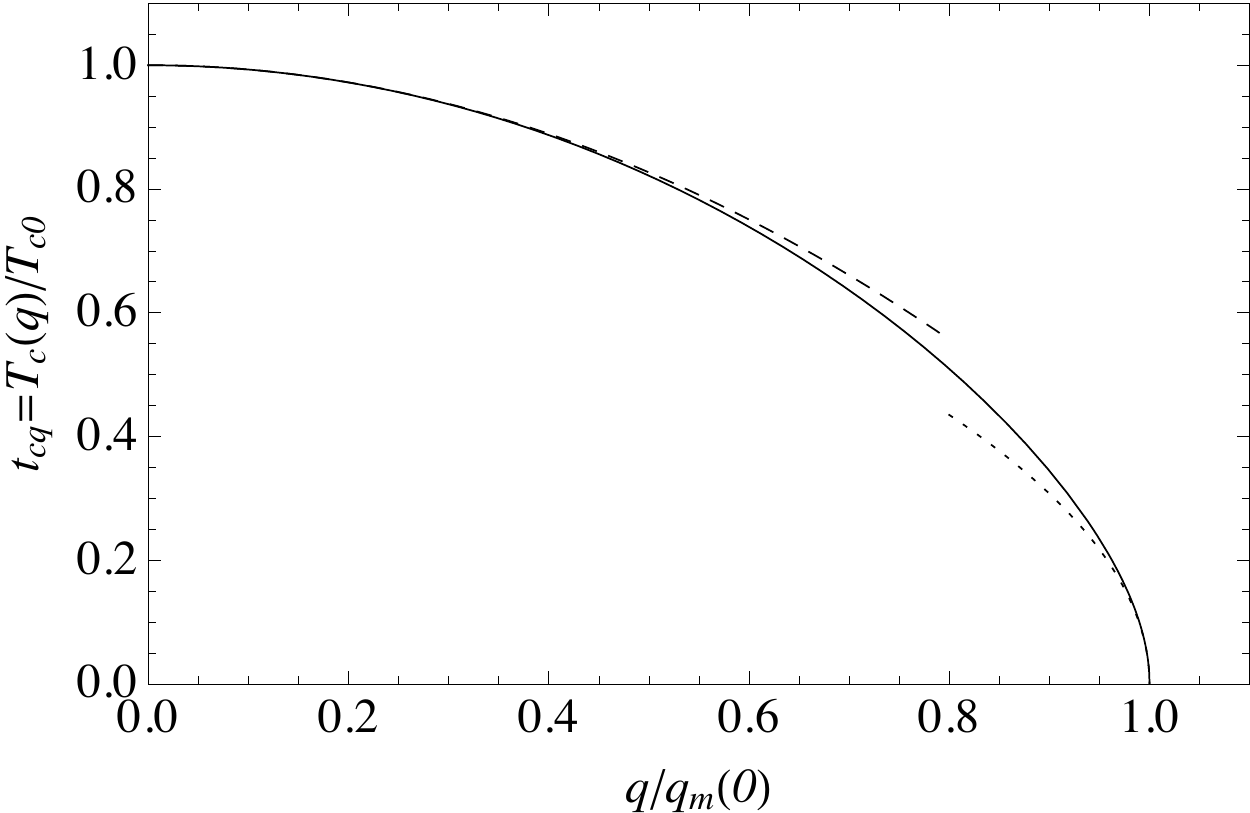}
\caption{%
$t_{cq} = T_{cq}/T_{c0}$ (solid) vs $q/q_m(0)$ obtained from Eqs.\ (\ref{deltaless})-(\ref{zetadef}). Also shown are expansions of $t_{cq}$ about $q=0$ [dashed, Eq.\ (\ref{tcq0})] and $q=q_m(0)$ [dotted, Eq.\ (\ref{tcq1})].  This figure is the same as a plot of $t=T/T_{c0}$ along the ordinate vs $q_m(T)/q_m(0)$ along the abscissa. $\Delta_q(T)>0$ only for values of $t$ and $q/q_m(0)$ under the curve. }
\label{tcqfig}
\end{figure}

To find the boundary in the $t$-$q$ plane where $\Delta_q(T)$ is reduced to zero, note from Eq.\ (\ref{unqomegaq}) that $\epsilon\sqrt{1+u_{nq}^2}\to n+1/2+Q/2\pi k_B T$ in the limit $\epsilon= \Delta_q(T)/2\pi k_B T \to 0$, such that Eq.\  (\ref{selfcons}) then yields the $q$-dependent transition temperature.   Defining $t_{cq} = T_{cq}/T_{c0}$, $P = Q/2\pi k_B T_{c0}$, $P_m = Q_m/2\pi k_B T_{c0}=e^{\psi(1/2)}=e^{-\gamma}/4 = 0.140$, $Q_m = \pi  k_B T_{c0} e^{-\gamma}/2 = \hbar D q_m^2(0)/2$, $q_m(0) = (\pi  k_B T_{c0} e^{-\gamma}/\hbar D)^{1/2} = (3/\pi\xi_0\ell)^{1/2}$, and  $v_m(0) = (\hbar/2m)(\pi  k_B T_{c0} e^{-\gamma}/\hbar D)^{1/2}$, we obtain
\begin{eqnarray}
  \ln \frac{1}{t_{cq}}&=& \sum_{n=0}^{\infty}\left(\frac{1}{n+1/2}- \frac{1 } { n+1/2+P/t_{cq}}\right),\, \label{tc1}\\
&=&\psi\left(\frac{1}{2}+\frac{P}{t_{cq}}\right)-
\psi\left(\frac{1 }{2}\right),
\label{tc2}
\end{eqnarray}
where $\psi$ is the digamma function and $P/P_m  = [q/q_m(0)]^2 = [v_s/v_m(0)]^2$.  Figure \ref{tcqfig} shows $t_{cq}$ as a function of $q/q_m(0)=v_s/v_m(0)$. Expansions of $t_{cq}$ about $q=0$ and $q = q_m(0)$ yield, respectively, the approximations
\begin{eqnarray}
t_{cq} &=&1-\frac{\pi^2 e^{-\gamma}}{8}\frac{q^2}{q_m^2(0)}=1-0.693\frac{q^2}{q_m^2(0)},\label{tcq0}\\
 &=&\sqrt{3}e^{-\gamma}\sqrt{1-\frac{q}{q_m(0)}}=0.972\sqrt{1-\frac{q}{q_m(0)}}, 
\label{tcq1}  
\end{eqnarray}
which are shown as the dashed and dotted curves in Fig.\ \ref{tcqfig}.

A similar procedure can be used to determine $q_m(T)$, the value of $q$ that drives $\Delta_q(T)$ to zero for a given value of $t$.  Equation (\ref{selfcons}) then yields 
\begin{eqnarray}
  \ln \frac{1}{t}&=& \sum_{n=0}^{\infty}\left(\frac{1}{n+1/2}- \frac{1 } { n+1/2+\alpha}\right),\, \label{alpham1}\\
&=&\psi\left(\frac{1}{2}+\alpha\right)-
\psi\left(\frac{1 }{2}\right),
\label{qmt}
\end{eqnarray}
where $\alpha$ is given in Eq.\ (\ref{alpham2}).
Since $t_{cq}$ and $q_m(T)$ are both determined by the equation obtained by setting $\Delta_q(T) =0$, it should not be surprising that a plot of $t$ vs  $q_m(T)/q_m(0)$ is exactly the same as the plot of $t_{cq}$ vs $q/q_m(0)$, shown in Fig.\ \ref{tcqfig}.
Expansions of $q_m(T)/q_m(0)$ about $t=0$ and $t=1$ yield, respectively, the approximations
\begin{eqnarray}
\frac{q_m(T)}{q_m(0)} &=&1-\frac{e^{2\gamma}}{3}t^2=1-1.057t^2,\label{qmt0}\\
 &=&\sqrt{\frac{8e^\gamma}{\pi^2}}e^{-\gamma}\sqrt{1-t}=1.202\sqrt{1-t}, 
\label{qmt1}  
\end{eqnarray}
which correspond to  the  dotted and dashed curves in Fig.\ \ref{tcqfig}.  Equations (\ref{qmt0}) and (\ref{qmt1}) are most easily obtained by making the replacements $q \to q_m(T)$ and $t_{cq} \to t$ in Eqs. (\ref{tcq1}) and (\ref{tcq0}).

The upper critical field is related to the temperature-dependent coherence length via $B_{c2}(T) = \phi_0/2\pi \xi^2(T)$, and, as discussed in Ref.\ \onlinecite{StJames69}, in the dirty limit $B_{c2}$ can be obtained from the equation
\begin{equation}
\ln\Big(\frac{1}{t}\Big) = \psi\left(\frac{1}{2}+\rho\right)-
\psi\left(\frac{1 }{2}\right),
\label{Bc21}
\end{equation}
where $t = T/T_{c0}$ and 
\begin{equation}
\rho = \Big(\frac{e^{-\gamma}}{4}\Big)\Big(\frac{B_{c2}(T)}{B_{c2}(0)}\Big)\frac{1}{t},
\label{Bc22}
\end{equation}
where $B_{c2}(0) = \phi_0/2\pi \xi^2(0)$ and  $\xi(0) = (\pi\xi_0\ell /3)^{1/2}$.
At $t=0$, we have $\lambda_0(0)/\xi(0) = \sqrt{3/\pi}\lambda_L(0)/\ell = 0.977 \lambda_L(0)/\ell$.  

As $t \to 0$, $B_{c2}(T)/B_{c2}(0) \approx 1-(2e^\gamma/3)t^2$.

As $t\to 1$, $B_{c2}(T)/B_{c2}(0) \to (8 e^{\gamma}/\pi^2)(1-t) = 1.444 (1-t)$.  Note that this result is consistent with what Helfand and Werthamer\cite{HW} found for their normalized field 
$h^*(0)=B_{c2}(0)/ (dB_{c2}/dt)_{t=1} =0.69=1/1.444$ in the dirty limit. Since $\xi(T) \to \sqrt{\pi^3e^{-\gamma}/24}\sqrt{\xi_0 \ell}/\sqrt{1-t}=0.852\sqrt{\xi_0 \ell}/\sqrt{1-t},$  we have in this limit $\lambda_0(T)/\xi(T) =\kappa = \sqrt{42 \zeta(3)/\pi^4}\lambda_L(0)/\ell = 0.720\lambda_L(0)/\ell$.

A plot of $B_{c2}(T)/B_{c2}(0)$ is equivalent to a plot of $[\xi(0)/\xi(T)]^2$, as shown in Fig.\ \ref{nsofig}.  Moreover, comparing Eqs.\ (\ref{Bc21}) and (\ref{Bc21}) with Eqs. (\ref{alpham1})-(\ref{alpham2}), we see that $q_m(T)=1/\xi(T)$ and $q_m(0)=1/\xi(0)$, and for all temperatures we have
\begin{equation}
\frac{B_{c2}(T)}{B_{c2}(0)}=\frac{\xi^2(0)}{\xi^2(T)}=\frac{q_m^2(T)}{q_m^2(0)}.
\label{Bc23}
\end{equation}

\section{Work done and free-energy changes resulting from current changes\label{EnergySec}}

It is of interest to examine the changes in energy as the current increases from zero to some final value.  The work done  per unit volume is
\begin{equation}
W_v=\int_0^{t}j_{sq'}E dt' = -\Big(\frac{\phi_0}{2\pi}\Big)\int_{0}^{q}j_{sq'}dq',
\label{Wv}
\end{equation}
which is equal to the change in the free-energy density, as we show below.

The free-energy density $\Omega_q(T) = F_S(T)-F_N(T)$ of a current-carrying superconductor relative to the energy density of the normal state can be obtained by taking advantage of the theoretical similarities to the problem of superconducting alloys containing paramagnetic impurities.   The expression for $\Omega_q(T) = F_S(T)-F_N(T)$ for the latter case obtained by Skalski et al.\cite{Skalski64} in their Eq.\ (5.6) can be rewritten compactly for the current-carrying superconductor as 
\begin{equation}
\Omega_q=N(0)\int_0^\infty {\rm Re}\Big[2p(\omega)+\frac{\Delta_q}{\sqrt{u^2-1}}\Big]\tanh\frac{\beta\omega}{2}d\omega,
\label{OmegaIntegral}
\end{equation}
using the replacements and changes in notation $\Gamma \to Q$, $\Delta(T,\Gamma) \to \Delta_q(T)$, $N_0 \to N(0)$, and $\omega_D' \to \infty$ (weak-coupling limit).  Here $\beta = 1/k_BT$, 
\begin{equation}
\frac{\omega}{\Delta_q} = u\left
(1-i\frac{\zeta}{ \sqrt{u^2-1}}\right), 
\label{uofomega}
\end{equation}
and 
\begin{eqnarray}
p(\omega)& =& -\int_\omega^\infty\Big[\frac{u'}{\sqrt{u'^2-1}}-1\Big]d\omega'\\
&=&\Delta_q\Big[\Big(1-i\frac{\zeta}{\sqrt{u^2-1}}\Big)(\sqrt{u^2-1}-u)\nonumber\\
&&-i\frac{\zeta}{2(u^2-1)}\Big]
\end{eqnarray}
is a quantity arising from partial integration of the term proportional to $\ln(1+e^{-\beta\omega})$ in the entropy contribution to $\Omega_q(T)$.  Some useful relations are Re[$p(0)$] = 0 and 
\begin{equation}
\int_0^\infty {\rm Re}\Big[p(\omega)+\omega\Big(\frac{u}{(u^2-1)^{1/2}}-1\Big)\Big] = -\frac{\Delta_q^2}{2}.
\end{equation}

The quantity within the brackets in the integrand of Eq.\ (\ref{OmegaIntegral}) can be expressed in terms of $u$  as 
\begin{equation}
\omega\Big(\frac{2\sqrt{u^2-1}}{u}-2+\frac{1}{u\sqrt{u^2-1}}\Big),
\end{equation}
and the integral in  Eq.\ (\ref{OmegaIntegral}) can be evaluated using contour integration around the boundaries of the first quadrant of the complex $\omega$ plane, taking into account the poles along the imaginary axis at the Matsubara frequencies $i\omega_n =i2\pi k_BT(n+1/2) =i2\pi k_BT\eta$.  The result is
\begin{eqnarray}
\!\!\!\Omega_q(T) &\!=\!&\!- N(0) (2\pi k_BT)^2\sum_{n=0}^\infty\eta\Big[2 \Big(\frac{u_{nq}}{\sqrt{1+u_{nq}^2}}-1\Big) \nonumber \\
&&+\frac{1}{u_{nq}\sqrt{1+u_{nq}^2}}\Big].
\label{OmegaSum}
\end{eqnarray}

Differention of Eqs.\ (\ref{selfcons}) and  (\ref{OmegaSum})
 with respect to $q$ yields the general result that 
\begin{eqnarray}
\frac{d\Omega_q(T)}{dq}&=& N(0) (2\pi k_BT)\frac{dQ}{dq}\sum_{n=0}^\infty\frac{1}{1+u_{nq}^2}
\label{dOmegaByq}\\
&=& -\Big(\frac{\phi_0}{2\pi}\Big)j_{sq}(T).
\end{eqnarray}
This result agrees with Eq.\ (\ref{Wv}).

In the zero-temperature limit, the sum in Eq.\  (\ref{OmegaSum}) can be converted to an integral over $\eta$ using Eq.\ (\ref{unqomegaq}), with the result\cite{Maki69}
\begin{eqnarray}
\!\!\!\!\!\Omega_q(0) &\!=\!&-\frac{N(0)\Delta_q(0)^2}{2}\Big(1-\frac{\pi \zeta_0}{2}+\frac{2 \zeta_0^2}{3}\Big),\;\zeta_0\le 1,\qquad
\label{Omegaq01}\\
 &\!\!=\!&-\frac{N(0)\Delta_q(0)^2}{2}\Big(1-\frac{\pi \zeta_0}{2}+\frac{2 \zeta_0^2}{3}\nonumber\\
&\!\!\!\!&-\frac{(2\zeta_0^2\!+\!1)\sqrt{\zeta_0^2\!-\!1}}{3\zeta_0}\!+\!\zeta_0\tan^{-1}\!\sqrt{\zeta_0^2\!-\!1}\Big),\nonumber\\
&\!\!\!&\;\;\;\;\;\;\;\;\;\;\;\;\;\;\;\;\;\;\;\;\;\;\;\;\;\;\;\;\;\;\;\;\;\;\;\;\;\;\;\;\;\;\;\;\;\;\;\;\;\;\;\zeta_0\ge 1,
\label{Omegaq02}
\end{eqnarray}
where $\Delta_q(0)$ is given by Eqs.\ (\ref{deltaless}) and (\ref{deltamore}), and   $\zeta_0$ by Eq.\ (\ref{zetadef}).

\begin{figure}
\includegraphics[width=8cm]{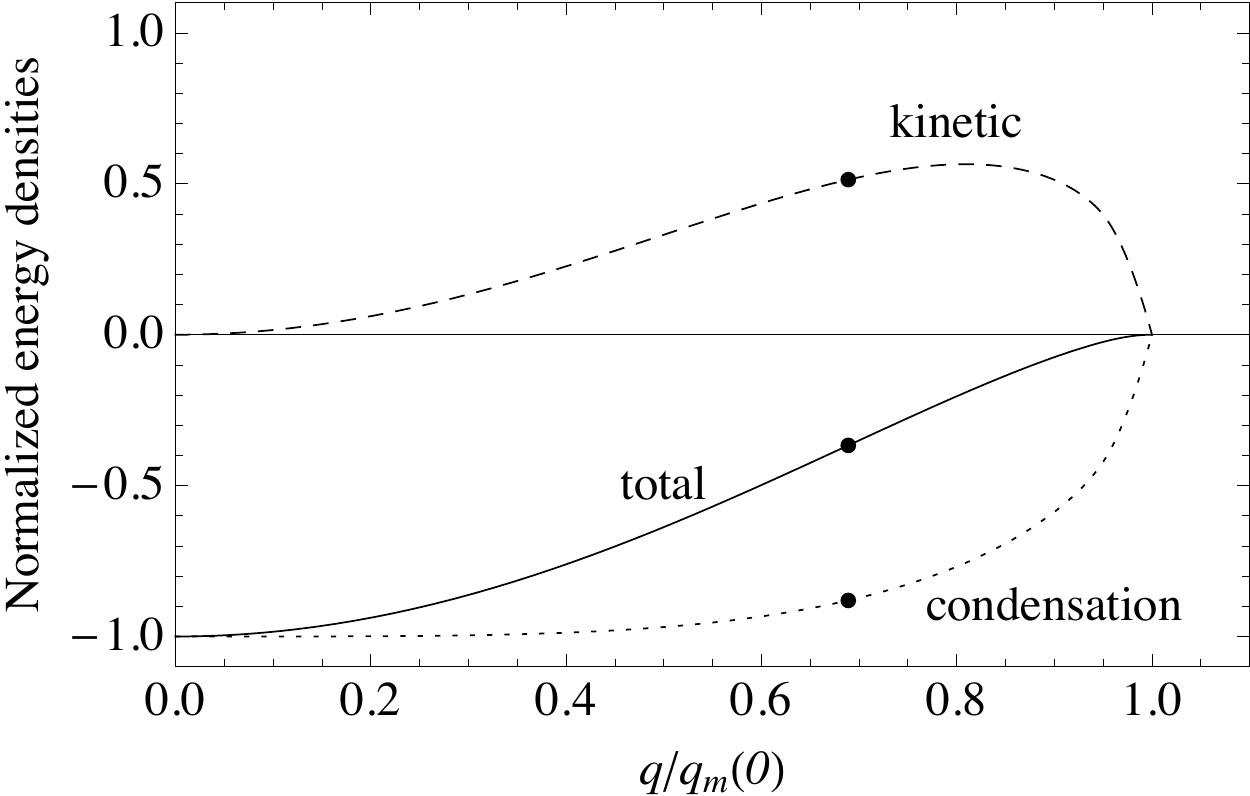}
\caption{%
Energy densities  at $T = 0$:  $\Omega_q(0)$ [solid curve, Eqs.\ (\ref{Omegaq01}) and (\ref{Omegaq02})], ${\cal F}_{cq}(0)$, [dotted, Eq.\ (\ref{Fcq})], and ${\cal F}_{kq}(0)$ [dashed,  Eq.\ (\ref{Fkq})],  normalized to $N(0)\Delta_0^2(0)/2$. The black points identify the values of these energy densities  at $q_d/q_m(0) = 0.689$, where $\Omega_q(0)$ has its maximum slope and the depairing current density is achieved.  }
\label{energiesfig}
\end{figure}
It is possible to write the free-energy density as the sum of two terms, $\Omega_q(T) = 
{\cal F}_{cq}(T) + {\cal F}_{kq}(T)$, where we identify ${\cal F}_{cq}(T)$ as the condensation-energy density and   ${\cal F}_{kq}(T)$ as the kinetic-energy density:
\begin{eqnarray}
\!\!\!{\cal F}_{cq}(T)\!&=&\!-N(0)(2\pi  k_BT)\sum_{n=0}^\infty\Big[2\hbar\omega_n \Big(\frac{u_{nq}}{\sqrt{1+u_{nq}^2}}-1\Big) \nonumber \\
\!\!\!\!&&+\frac{\Delta_q}{\sqrt{1+u_{nq}^2}}\Big],
\label{Fcq}\\
\!\!\!{\cal F}_{kq}(T)\!&=&\!N(0) (2\pi k_BT)Q\sum_{n=0}^\infty\frac{1}{1+u_{nq}^2}.
\label{Fkq}
\end{eqnarray}
Note, however, that both terms depend upon $q$ and interact.  As $q$ increases from zero, 
${\cal F}_{kq}(T)$ initially increases from zero and  the magnitude of ${\cal F}_{cq}(T)$ decreases, but their sum $\Omega_q(T)$ decreases to zero as $q \to q_m(T)$.  Figure \ref{energiesfig} shows the $q$ dependence of the energy densities  $\Omega_q$, ${\cal F}_{cq}$, and ${\cal F}_{kq}$ at $T = 0$.

It is important to note that ${\cal F}_{kq}(T)$ is {\it not} equal to ${\cal L}_{k}(q,T) j_{sq}^2/2$ except in the limit as $q\to 0$ [see Eq.\ (\ref{uk})].

The bulk thermodynamic critical field $H_c(T)$ is defined via the superconducting condensation energy at $q$ = 0:
\begin{eqnarray}
\frac{1}{2}\mu_0H_c^2(T)&=& -\Omega_0(T) \\
&=&N(0)(2\pi k_BT)^2\sum_{n=0}^\infty\Big(2\sqrt{\eta^2+\epsilon_0^2}\nonumber\\
&&-2\eta -\frac{\epsilon_0^2}{\sqrt{\eta^2+\epsilon_0^2}}\Big),
\label{HcT}
\end{eqnarray}
where $\eta = n+1/2$ and $\epsilon_0 = \Delta_0(T)/2\pi k_B T$. Equation (\ref{HcT}), which follows from Eq.\ (\ref{OmegaSum}) when $q \to 0$ and  $u_{nq} \to u_{n0} = \eta/\epsilon_0$ [see Eq.\  (\ref{unqomegaq})], reproduces the BCS\cite{BCS57} temperature dependence of $H_c(T)$.


\begin{thebibliography}{99}
\bibitem{foot1} Here we discuss the flow of electrons, but instead we could have discussed the flow of Cooper pairs of density $n_s^*=n_s/2$, charge magnitude $e^* = 2e$, and mass $m^* = 2m$.  This would have yielded expressions such as $\mu_0 \lambda_{0}^2(T)= m^*/n_{s0}^*(T)e^{*2}$ and $j_{sq} = -n_{sq}^* e^* v_s$.
\bibitem{foot2} When $\nabla \cdot \bm j_s=0,$ the conditions $P = 0$ and $E= -dA_s/dt$ hold  within the generalized time-dependent GL theory,\cite{Kopnin01} and we assume that these conditions also apply here.
\bibitem{foot3} This definition of the inductivity is the natural extension of the definition of the inductance $L$ via  $V = LdI/dt$ relating the voltage drop $V$ across an inductor to the current $I$.
\end{thebibliography}
\end{document}